%% file: main.tex
\documentclass[%
 reprint,
 amsmath,amphysicalsymb,
 aps,
pra,
]{revtex4-1}

\input{preamble}

\begin{document}

\preprint{APS/123-QED}

\title{
 Achieving computational gains with quantum error-correction primitives:\\Generation of long-range entanglement enhanced by error detection
}

\author{Haoran Liao}
\email{haoran.liao@q-ctrl.com}
\author{Gavin S. Hartnett}
\author{Ashish Kakkar}
\author{Adrian Tan}
\author{Michael Hush}
\author{Pranav S. Mundada}
\author{Michael J. Biercuk}
\author{Yuval Baum}
\affiliation{Q-CTRL, Los Angeles, CA USA and Sydney, NSW Australia}

\begin{abstract}
The resource overhead required to achieve net computational benefits from quantum error correction (QEC) limits its utility while current systems remain constrained in size, despite exceptional progress in experimental demonstrations. In this paper, we demonstrate that the strategic application of QEC primitives without logical encoding can yield significant advantages on superconducting processors---relative to any alternative error-reduction strategy---while only requiring a modest overhead. 
We first present a novel protocol for implementing long-range CNOT gates that relies on a unitarily prepared Greenberger-Horne-Zeilinger (GHZ) state as well as a unitary disentangling step; the protocol natively introduces an error-detection process using the disentangled qubits as flags. 
We demonstrate that it achieves state-of-the-art gate fidelities of over 85\% across up to 40 lattice sites, significantly and consistently outperforming the best alternative measurement-based protocol without introducing any additional ancilla qubits.
We then apply sparse stabilizer measurements to generate large GHZ states by detecting bit-flip and amplitude-damping errors. Employing this technique in combination with deterministic error suppression, we generate a 75-qubit GHZ state exhibiting genuine multipartite entanglement, the largest reported to date. The generation requires no more than 9 ancilla qubits and the fraction of samples discarded due to errors grows no higher than 78\%, far lower than previous discard fractions required for tests using comparable numbers of fully encoded qubits. 
This work in total represents compelling evidence that adopting QEC primitives on current-generation devices can deliver substantial net benefits.

\end{abstract}

\maketitle

\section{Introduction} 
Progress in experimental quantum error correction (QEC) has been rapid~\cite{Krinner2022, Zhao2022, google2023suppressing, gupta2024encoding, bluvstein2024logical, acharya2024quantum, reichardt2024tesseract, matsos2024, lachancequirion2023, reichardt2024atomcomputing}, with recent demonstrations achieving performance beyond break even, where QEC provides a net improvement in error rates over bare hardware. However, using QEC to achieve fault-tolerant computation still requires significant overheads~\cite{Babbush2021}, with estimates of hundreds or thousands of physical qubits per logical qubit required to achieve error rates compatible with large-scale algorithms~\cite{ Litinski2019gameofsurfacecodes}.  In response, so-called early-fault-tolerance frameworks have been proposed~\cite{dutkiewicz2024earlyft}, where overheads are reduced by targeting noise-resilient algorithms and achievement of less stringent logical-error rates. 
Even with these compromises, experimental demonstrations remain a challenge, and full QEC implementations on current-era devices typically deliver modest benefits at high computational costs and a large overhead of physical qubits.  Similarly, demonstrations of post-selected error detection on experiments using a few dozen encoded logical qubits can have exceedingly high discard rates above 99.9\%~\cite{bluvstein2024logical}, posing another significant form of overhead for users with constrained access to hardware (by e.g., cost or time).

In the interim, the community has developed a range of error-reduction techniques that do not require full QEC and its associated overheads. This begins with hardware-level innovation where progress in qubit coherence and gate errors has been substantial~\cite{Sundaresan2020, Carroll2022, Bal2024, Sete2024}. To address the inevitable residual impact of noise encountered in even the most advanced devices, several quantum error mitigation approaches have been proposed as a near-term strategy for addressing readout~\cite{mthree, mundada2023experimental} and computational errors~\cite{Kandala2019, Huggins_2021, van_den_Berg_2023, Kim2023, Liao2023}. These use post-processing of noise-corrupted quantum-computer outputs to improve the estimation of expectation values; this ensemble-post-processing approach is unfortunately largely incompatible with the fundamental concept of runtime error identification and correction underpinning QEC. In parallel, there has been rapid development of QEC-compatible routines based on gate and circuit-level error suppression~\cite{mundada2023experimental, coote2024, Seif2024, Tripathi2022, Pokharel2018, carvalho2021, Nation2023, hartnett2024}.  These approaches suppress errors at runtime and incur no overhead in circuit execution, enabling successful demonstrations of quantum algorithms at scales exceeding 100 qubits~\cite{sachdeva2024, Kim2023, bluvstein2024logical, Liao2023}.  With current modestly sized devices, users generally achieve greater overall quantum computational capability by selecting these alternative strategies over QEC, in order to reduce error rates while preserving larger effective qubit counts. Nonetheless,  these techniques cannot address all possible errors, with particular shortcomings in addressing (Markovian) incoherent computational errors such as pure dephasing or bit-flip errors.

In this work, we demonstrate that net computational improvements can be achieved on current-generation quantum computers using low-overhead QEC primitives~\footnote{By QEC primitives, we refer to core components or sub-routines employed in QEC, such as encoding, parity checks, non-destructive measurements via ancillas, and decoding.}---the underlying protocols employed in QEC---on the physical level without the need for QEC encoding.  We employ ancilla qubits and entangling parity checks (core elements of all stabilizer QEC codes~\cite{Nielsen_Chuang2011}) to perform error detection on unencoded computational qubits. We present two key demonstrations leveraging error-detection strategies to deliver enhanced overall capability.  First, we improve the implementation of a long-range CNOT gate using a novel protocol based on unitary preparation of a Greenberger-Horne-Zeilinger (GHZ) state followed by a predominantly unitary disentangling step. This approach has the advantage that the final state of the disentangled qubits  (the intermediate qubits that connect the distant control and target qubits themselves) reveals errors that have occurred during the gate.  With this approach we perform a long-range CNOT achieving over 85\% fidelity across up to 40 lattice sites, outperforming state-of-the-art alternatives on superconducting processors. Second, we generate large GHZ states using a protocol first presented by Mooney {\em et al.}~\cite{mooney2021generation}, which allows for the integration of sparse error detection through ancillary stabilizer measurements.  Using this resource-efficient routine (consuming no more than 9 flag qubits) in combination with a deterministic error-suppression pipeline~\cite{mundada2023experimental, coote2024}, we achieve genuine multipartite entanglement for up to 75 qubits, verified in terms of multiple-quantum coherence (MQC) fidelity---a record in the published literature.  In contrast to the work of Mooney {\em et al.}, we see significant enhancements in the fidelity of the GHZ state through the addition of error detection with increasing numbers of flags. We also observe a comparatively low discard rate, where over $80\%$ of the shots are kept in the case of generating a 27-qubit GHZ state and over $21\%$ in the 75-qubit state. These results demonstrate that incorporating QEC primitives on the physical level can deliver a substantial net improvement in the capability of a near-term quantum computer relative to the best alternative strategies. 

The remainder of this paper is structured as follows: 
In Sec.~\ref{sec:cnot}, we present the design of a new protocol for implementing the long-range CNOT gate.  This is followed by experimental demonstrations in Sec.~\ref{sec:teleported_cnot_results}; there we show significant improvement in a teleported CNOT on an IBM superconducting processor compared to the alternative measurement-based protocol.  We present both shot-by-shot experiments and experiments incorporating post-processing to implement readout error mitigation.
In Sec.~\ref{sec:ghz}, we introduce the method by which large-scale, high-fidelity GHZ states can be generated, presenting the principles behind sparse parity checks. Subsequently, in Sec.~\ref{sec:2d_ghz_results}, we experimentally demonstrate the method, characterizing performance via both the MQC and Hellinger fidelities. Our presentation includes detailed error analysis, highlighting performance variability with the number of flag qubits, and the role of error suppression in circuit execution.

\section{Long-range CNOT with error detection} \label{sec:cnot}

\begin{figure}[!b]
\centering
\includegraphics[width=0.48\textwidth]{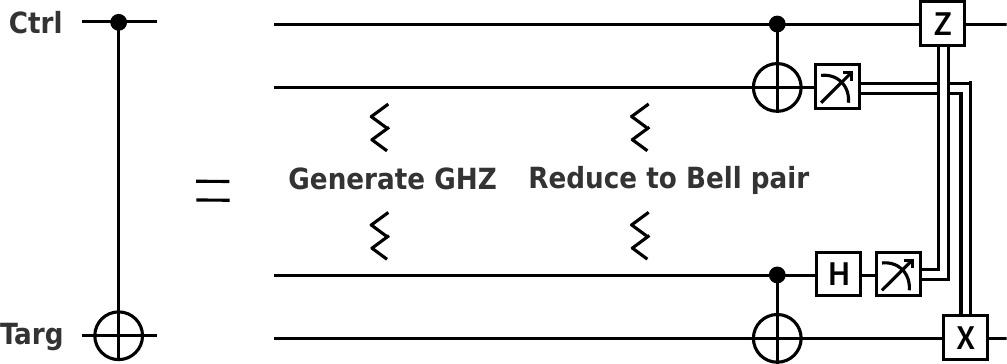}
\caption{ 
	\textbf{Quantum circuit for a general three-step teleportation-based framework for long-range CNOT.} Single lines indicate qubits, while double lines represent the flow of classical bits. Zigzag lines denote intermediate qubits, excluding the two qubits adjacent to the control and target qubits, which form a Bell pair.
	}
\label{fig:cnot_framework}
\end{figure}

\begin{figure*}[!ht]
\centering
\includegraphics[width=0.92\textwidth]{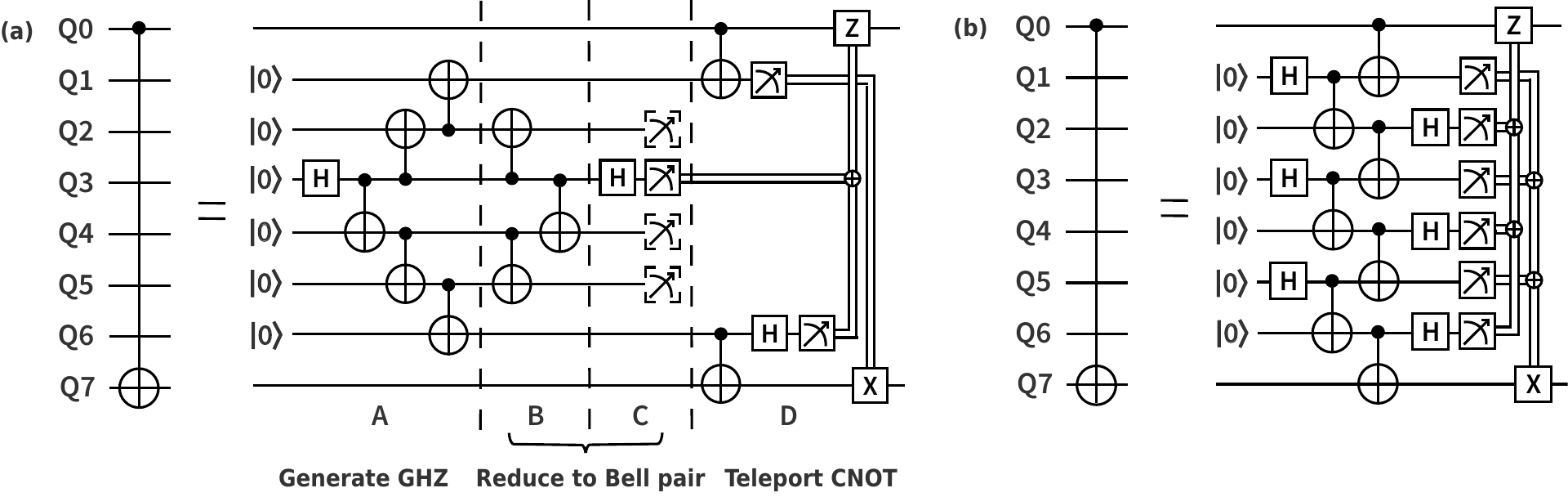}
\caption{
\textbf{
Two teleportation-based long-range CNOT protocols.} In both protocols, the initial states of the control and target qubits are arbitrary, whereas the intermediate qubits are initialized to the ground state. 
(a) Unitary entangle-disentangle protocol. Measurements with a dashed border are optional and required only for error detection. The protocol steps are separated by vertical dashed lines and indicated with the letters A, B, C, and D.
Here and in all subsequent figures, the $\oplus$ symbol applied to classical bits represents addition modulo 2. 
(b) Measurement-based protocol of Ref.~\cite{baumer2024efficient}.
}
\label{fig:teleported_cnot}
\end{figure*}

Improving the ability to achieve high-fidelity long-range CNOT gates is highly desirable in solid-state devices, where physical circuit architectures constrain effective device connectivity. These gates are particularly important for the large-scale implementation of algorithms like the quantum Fourier transform and fermionic simulations~\cite{Holmes_2020}, and facilitate the use of non-local QEC codes such as quantum low-density parity-check (qLDPC) codes~\cite{bravyi2024high, LDPC_2024}.
Previous work by B\"{a}umer {\em et al.}~\cite{baumer2024efficient} demonstrated teleportation of a CNOT gate using a measurement-based protocol across up to 100 lattice sites, achieving the highest fidelities for such tasks on superconducting processors at the time. 
In this section, we present a novel protocol for long-range CNOT gates that achieves superior performance by relying primarily on unitary operations and incorporating error detection.

\subsection{Novel circuit for long-range CNOTs}
As a generic matter, long-range CNOTs can be implemented through a three-step framework, as depicted schematically in Fig.~\ref{fig:cnot_framework}. First, one generates the shortest-path linear-chain GHZ state connecting the control and the target qubits. In the computational basis, an $n$-qubit ($n>2$) GHZ state is represented
\begin{equation}
    |\psi_{\text{GHZ},n}\rangle = \frac{1}{\sqrt{2}} \left( \ket{0}^{\otimes n}+\ket{1}^{\otimes n} \right).
\end{equation}
Second, one reduces the GHZ state to a Bell pair,  with each member of the pair neighboring the control and the target qubits.
In the third and final step, local operations and classical communication (LOCC) based on the Bell pair are used to implement the logic of the CNOT gate between the control and target qubits~\cite{Gottesman1999, Eisert_2000, circuit_knitting, vazquez2024combineprocessors}. By modifying the LOCC operations, this framework can be generalized to realize any controlled-unitary gate~\cite{Eisert_2000}. The protocol described above is general; concrete implementations can vary significantly based on details of the GHZ-state-generation and Bell-pair-reduction steps. 

Our new protocol follows this generalized three-step framework, and is depicted in Fig.~\ref{fig:teleported_cnot}(a) for the case of two qubits (Q0 and Q7) separated by six intermediate qubits (Q1--Q6).  In step A,  a linear-chain GHZ state is produced along the shortest path (Q1--Q6) between the target and control qubits, through repeated application of CNOT gates between adjacent qubits. The GHZ state is then reduced to a Bell pair on Q1 and Q6 through a combination of unitary operations and measurements in steps B and C. In particular, in step B, all intermediate qubits (forming a GHZ state) except one are unitarily disentangled through the action of local CNOT gates, leaving behind a smaller GHZ state on Q1, Q3, and Q6. In step C, the last remaining GHZ data qubit Q3 is projectively disentangled through an $X$-basis measurement; its $X$-basis measurement outcome determines the parity of the remaining Bell pair on Q1 and Q6, hence the feedforward $Z$ gate. 
Other unitarily disentangled qubits (Q2, Q4, and Q5) can be optionally measured and used as flag qubits for error detection, which we explain in detail below. Lastly, step D proceeds as the LOCC step of the general framework discussed above to achieve the CNOT gate. In particular, we design both the generation and the reduction schemes (steps A to C) to rely primarily on unitary operations as opposed to projective measurements, better suiting the relative error budgets on current devices. We refer to our new protocol as the ``unitary entangle-disentangle'' protocol. 

The disentangled qubits serve different purposes at different stages of the algorithm: as entanglement mediators during the creation of the long-range Bell pair state, and as flag qubits at the conclusion of the protocol. One can gain insight into this protocol by noting that the local CNOT gates in step B rotate the $n$-qubit GHZ state of step A into a product of a 3-qubit GHZ and the ground state, independently of the state of the control and target qubits. Concretely, the state of the intermediate qubits after step B is:
\begin{equation} \label{eq:intermediate_state_for_CNOT}
	\frac{1}{\sqrt{2}} \left( \ket{000} + \ket{111} \right) \otimes \ket{0}^{\otimes (n-3)} \,.
\end{equation}
Among the qubits in the GHZ state, the first and last are the neighbors of the control and target qubits, while the second lies at the halfway point of the teleportation path (rounded up or down by one lattice site if $n$ is even). 
This structure allows the disentangled qubits to serve as flag qubits; any deviation in readout on these $n-3$ qubits from the all-0 bit-string indicates the occurrence of an error. Bit-flip and amplitude damping errors that occur during the generation of the GHZ step or the reduction to Bell pair step may be captured, but due to the circuit structure, only $Z$-type errors that occur in the Hadamard gate in step A of the circuit in Fig.~\ref{fig:teleported_cnot}(a) are detectable. Detector theory~\cite{McEwen_2023} can also be applied here to identify the locations of the errors.

\begin{figure*}[!ht]
\centering
\includegraphics[width=0.87\textwidth]{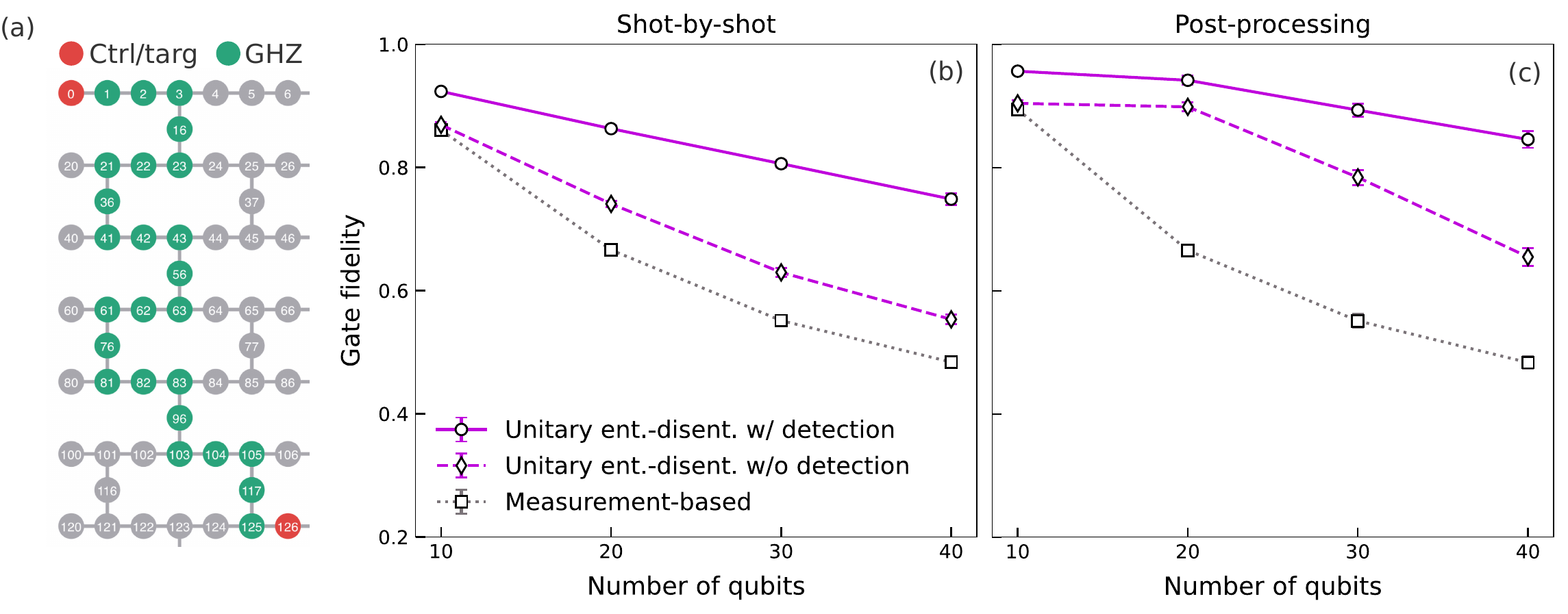}
\caption{
\textbf{Experimental results for long-range CNOT protocols.} 
(a) The geometrical extent of the long-range CNOT, showing the control and target qubits (red) connected via a shortest path of 25 intermediate qubits (green), which are prepared in a GHZ state during step A of the protocol.
(b, c) Comparison of the average gate fidelity between the unitary entangle-disentangle (unitary ent.-disent.) protocol and the measurement-based protocol, as a function of teleportation distance (measured in qubits): (b) fidelity from shot-by-shot experiments and (c) fidelity after post-processing with additional readout error mitigation. 
Although both protocols feature conditional feedforward operations based on measurement outcomes, for simplicity in implementation, shots requiring feedforward were discarded, reducing the overall yield---though in all cases, a minimum remaining shot count of $1{,}000$ was ensured after overall post-selection.
For the unitary entangle-disentangle protocol with error detection, erroneous shots due to the activation of flag qubit(s) were discarded. Data were collected consecutively on \texttt{ibm\_fez} to ensure direct comparability without concern over hardware drift. Error bars represent $\pm 1$ standard deviation from $50$ bootstraps applied to a single experimental run.
}
\label{fig:teleported_cnot_results}
\end{figure*}

This new long-range CNOT gate protocol can be compared to existing protocols in order to benchmark performance.  A baseline protocol uses SWAP gates to permute the control and target qubits to be adjacent, after which a local CNOT gate is applied, and, if necessary, the SWAP-induced permutation is undone. Recently, B\"{a}umer {\em et al.}~\cite{baumer2024efficient} demonstrated a constant-depth measurement-based protocol, depicted in Fig.~\ref{fig:teleported_cnot}(b), which consistently outperformed the baseline SWAP-based construction. In fact, this protocol conforms to the generalized three-step (GHZ $\rightarrow$ Bell pair $\rightarrow$ gate teleportation)  framework, which we show explicitly through an alternative derivation of the circuit in App.~\ref{app:extended_derivation_dynamic_cnot}.  In that derivation, we explain that low-overhead error detection is not available for the measurement-based protocol. Finally, in App.~\ref{app:fan_out}, we provide a generalization of the unitary entangle-disentangle protocol for the long-range quantum fan-out gates~\cite{baumer2024_more_gates, Paul2013, Hoyer2005, allcock2023, Xu_2023}.

\subsection{Experimental results} \label{sec:teleported_cnot_results}
We implement the novel unitary entangle-disentangle protocol for long-range CNOT generation on an IBM quantum computer; an example of the implementation mapped to a physical-device layout is depicted in Fig.~\ref{fig:teleported_cnot_results}(a), where the protocol teleports a CNOT gate between the control and target qubits (red) along a linear chain of qubits (green). Errors can be detected by measuring the unitarily disentangled GHZ qubits (which effectively serve as flags), and erroneous samples (shots) are discarded if any of the flags is measured to be in $\ket{1}$. Hereafter, we use a ``Q'' notation indicating the system size, measured in the number of qubits. 

In Fig.~\ref{fig:teleported_cnot_results}(b) and (c) we present measurements of the effective gate fidelity as a function of system size.  
We determine the average gate fidelity using Monte Carlo process certification~\cite{flammia2011direct, da_Silva_2011}, following the procedure described in Ref.~\cite{da_Silva_2011, baumer2024efficient}. The different panels present both a shot-by-shot mode, where each sample is considered in isolation, and a post-processing mode where multiple shots are aggregated statistically before the application of readout error mitigation as in Ref.~\cite{mundada2023experimental} on all bits, followed by post-selection on the intermediate qubits. For each measurement approach, we separately present data with and without the use of error detection and compare against the state-of-the-art measurement-based protocol~\cite{baumer2024efficient}.  We do not experimentally implement the SWAP-based protocol due to its known inferior performance.  All experiments incorporate error suppression in circuit execution~\cite{mundada2023experimental, coote2024} to establish a high-fidelity baseline against which the addition of error detection can be evaluated in isolation.

In both modes, the unitary entangle-disentangle protocol achieves consistently and significantly higher average gate fidelity than the measurement-based protocol for teleportation across up to at least $40$ qubits (Figs.~\ref{fig:teleported_cnot_results}(b) and (c)). We achieve $\gtrsim 90\%$ average gate fidelity up to 20Q in shot-by-shot experiments and up to 30Q when incorporating readout error mitigation in post-processing. Experiments with and without error detection both outperform the benchmark, demonstrating the relevance of the protocol design to performance.  Taking advantage of the in-built error detection protocol reduces the effective gate infidelity by around $50\%$. The highest performance is achieved when also incorporating readout error mitigation in post-processing, consistent with the known error budgets for currently available devices, in which gate errors are lower than readout errors. It is important to emphasize the reasonably low runtime overhead due to the error detection of our protocol---for these experiments, the discarded fraction (fraction of experiments identified as failed in the error-detection routine)  grows to at most approximately $80\%$ for the largest gate teleportation lengths in shot-by-shot mode: see App.~\ref{app:cnot}. There, we also provide an expanded comparison between the unitary entangle-disentangle protocol and the measurement-based protocol.    

\section{GHZ state preparation with error detection}
\label{sec:ghz}

GHZ states are maximally entangled multi-qubit states that carry significant roles in the broad field of quantum information.  This includes use in entanglement-enhanced quantum sensing~\cite{quantum_sensing_review2017}, QEC, and multi-party quantum communications~\cite{Hillery1999SecretSharing}. The fidelity with which these states can be prepared in a quantum computer has also been proposed as a performance benchmark for quantum hardware~\cite{wei2020verifying, Kam2024}. These varied applications have led to substantial efforts attempting to generate large GHZ states on state-of-the-art quantum computers~\cite{mooney2021generation, baumer2024efficient}, involving both conventional unitary circuit and measurement-based preparation relying on mid-circuit measurement and feedforward (dynamic circuits)~\cite{Andersen_2020, Bo2022, Moses_2023, chen2023nishimori, baumer2024efficient, hashim2024}.  

The $n$-qubit GHZ state has an extreme sensitivity to noise---a sensitivity which grows with the number of entangled qubits~\cite{Monz2011}.  It is our objective to deploy QEC primitives in order to enhance the maximum circuit width capable of exhibiting genuine multipartite entanglement in the presence of realistic noise encountered in a quantum computer.

We adopt the strategy of unitary preparation and incorporation of sparse parity checks introduced by Mooney {\em et al.}~\cite{mooney2021generation}, combining this technique with implementation of an error-suppression pipeline for arbitrary circuit execution~\cite{mundada2023experimental, coote2024}.  This general approach has the advantage of avoiding the penalties associated with readout errors in state preparation (see App.~\ref{app:bitstring_spreading} for details on the inherent ineffectiveness of readout error mitigation for the measurement-based state-preparation strategy). As the implementation details of the error-suppression protocols have been published previously, we reserve discussion here to cover the relevant QEC primitives, and only include analysis of the impact of error suppression in our experimental measurements.  

The use of parity checks for error detection exploits the fact that an $n$-qubit GHZ state is a stabilizer state with
$2^n$ stabilizers generated by
\begin{align}\label{eq:ghz_stabilizer}
\langle Z_1Z_2,\ Z_2Z_3,\ \dots,\ Z_{n-1}Z_{n},\ X_1X_2\cdots X_n \rangle \,.
\end{align} 
As the GHZ state is uniquely determined by its stabilizers, any infidelity in the state is manifest as a violation of the stabilizer condition for one or more of the stabilizers. Hence, measuring such violations enables direct error detection---we again consider this to be a QEC primitive designed to reduce the impact of errors without the need for full QEC encoding. 

In our implementation, a selected set of stabilizers of the form $Z_i Z_j$ (each can be expressed as a product of some stabilizer generators defined in Eq.~\eqref{eq:ghz_stabilizer}) is measured non-destructively by the use of a small number of additional ancilla qubits (flags).  We classify the possible parity checks as type-$s$, where $s$ indicates the number of additional SWAP gates required for the error-detecting stabilizer measurement. 

\begin{figure*}[!ht]
\centering
\includegraphics[width=1\textwidth]{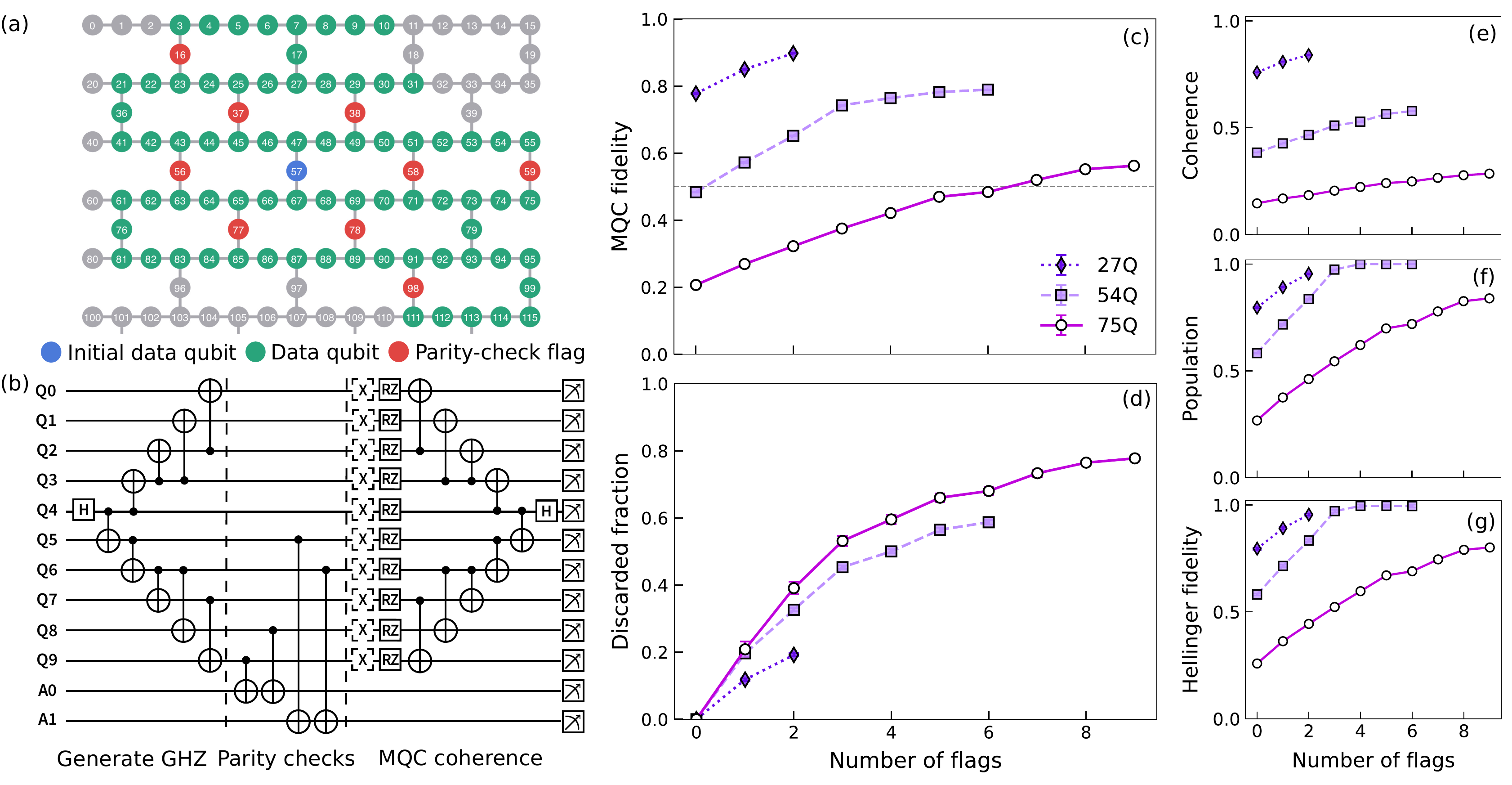}
\caption{\textbf{Generation of GHZ states with sparse parity checks, followed by multiple-quantum coherence (MQC) verification of the states.} 
(a) The layout of the generated $75$Q GHZ state. The state is generated by a unitary circuit starting at the initial data qubit (blue), with entanglement grown in a breadth-first-search manner to include other data qubits (green). Minimal-overhead (type-0) ancilla qubits (flags) are shown in red. 
(b) An example circuit generating a GHZ state on Q0--Q9. Two type-0 flags A0 and A1 check the parities of (Q8, Q9), and (Q5, Q6), respectively. The circuit (labeled MQC coherence) is appended to measure the MQC fidelity of the generated GHZ state. 
(c) The MQC fidelity as a function of the number of type-0 flags used in post-selection, evaluated on IBM 156Q Heron device \texttt{ibm\_fez}; (d) the discarded fraction of shots averaged over circuits---in particular, the discarded fraction with all $9$ flags on 75Q GHZ preparation is $78(1)\%$. 
(e) The coherence, (f) population, and (g) Hellinger fidelity, also as a function of the number of flags. 
Circuits are sampled at $10{,}000$ shots for 27Q and 54Q, and at $35{,}000$ shots for 75Q. Error bars show $\pm 1$ standard error, most of which are not visible (error bars show shot noise, and not fluctuations of the device over time).
}
\label{fig:layout_and_gen_circ}
\end{figure*}

\begin{figure}[!t]
\centering
\includegraphics[width=0.9\columnwidth]{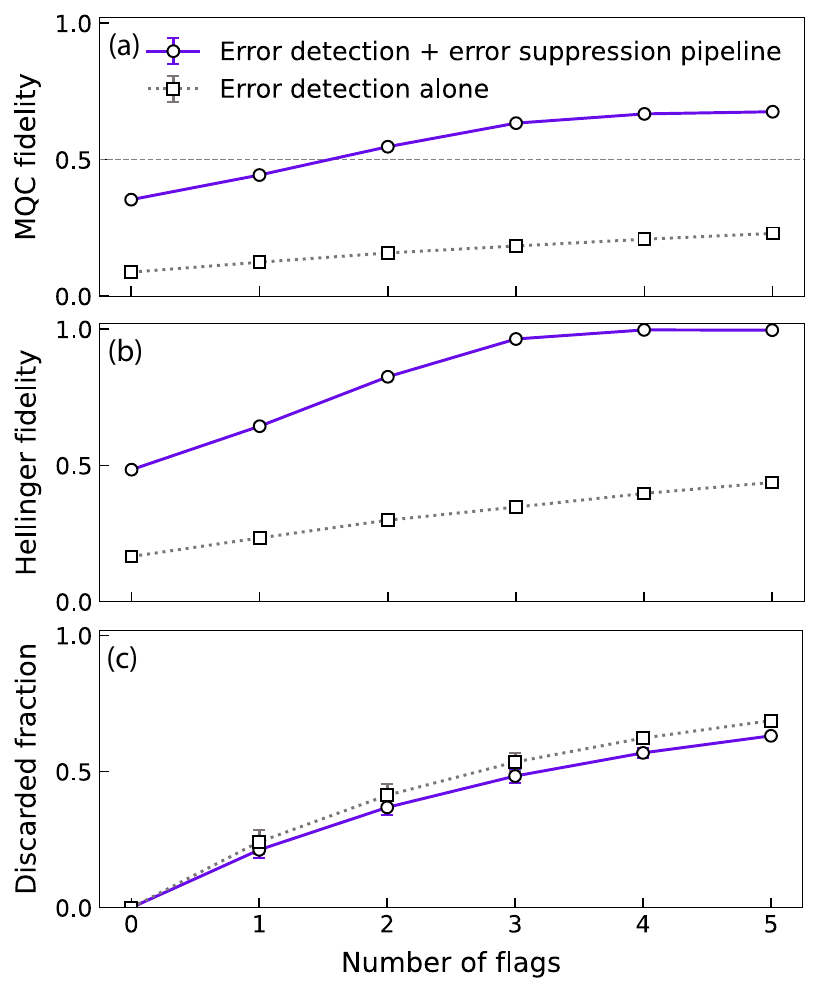}
\caption{
\textbf{Comparison between GHZ-state-generation performance with and without inclusion of the error suppression pipeline.} (a) The MQC fidelity, (b) Hellinger fidelity, and (c) discarded fraction of shots as a function of the number of type-0 flags evaluated for the generation of a 45Q GHZ state on \texttt{ibm\_brisbane}, with and without error suppression. The horizontal dashed line at an MQC fidelity of $0.5$ indicates the threshold of genuine multipartite entanglement. Circuits are sampled with $10{,}000$ shots, and the discarded fraction of shots is averaged over circuits. The parts of the error suppression pipeline are detailed in Ref.~\cite{mundada2023experimental}. Error bars show $\pm 1$ standard error, most of which are not visible.
}
\label{fig:against_qiskit_45q}
\end{figure}

The positioning of the flag qubits is not unique and is influenced both by the operational overhead of the state preparation procedure and the device topology.  On a heavy-hex topology~\cite{Chamberland2019}, an efficient placement strategy includes placing the flag qubits on the weak links (the low-density rows), and keeping enough weak links ``free'' to allow for efficient (layer-wise) expansion of the GHZ state. An example of a physical layout incorporating a set of minimal-overhead type-0 parity-check flag qubits is shown in Fig.~\ref{fig:layout_and_gen_circ}(a).  This configuration adds 9 flag qubits to a GHZ state with 75 data qubits, such that no SWAP gates are needed for either the state generation or the parity checks.  In the resulting circuit, there are $43$ two-qubit-gate layers, just one larger than the $42$ layers required for an equivalent state-generation protocol with no flags.

\subsection{State-preparation quality measures} \label{sec:2d_ghz_results}
We employ a commonly used characterization method, the multiple-quantum coherence (MQC)~\cite{wei2020verifying, mooney2021generation, Bao2024}, to validate the quality of the GHZ states generated in the experiments described below. We also consider the Hellinger fidelity as a complementary quality measure which is lighter weight in implementation.

MQC exploits the structure of the GHZ state to measure the fidelity with a number of experiments that scales linearly in system size $n$. Importantly, to establish whether a state exhibits genuine multipartite entanglement, it is sufficient to show MQC fidelity $F > 0.5$~\cite{leibfried2005creation}. Here we briefly describe the two quantities measured in MQC: the coherence $C$, and the population $P$.  Once obtained, the MQC fidelity $F$ is the arithmetic mean of the coherence $C$ and population $P$.  Further details on MQC are provided in App.~\ref{app:ghz}. 

The coherence $C \in [0,1]$ verifies the off diagonals of the density matrix of a prepared GHZ state. The main idea is to measure the interference pattern (parity oscillation) of a phase-rotated GHZ state, i.e., ${\left(\ket{00...0}+e^{in\phi}\ket{11...1}\right)/\sqrt{2}}$. 
Experimentally one measures the interference pattern between the prepared non-phase-rotated and phase-rotated states, $\rho$ and $\rho_{\phi}$, respectively, defined as ${S_{\phi} := \text{Tr}(\rho \, \rho_{\phi})}$. This quantity should ideally exhibit a sinusoidal behavior as a function of $\phi$ (see Fig.~\ref{fig:layout_and_gen_circ}(b) for the corresponding measurement circuit).  
Specifically, the coherence $C$ is related to the $n$-th real Fourier mode, $I_n$, of $S_{\phi}$ as
 \begin{equation}
     C := 2\sqrt{I_n}.
 \end{equation} 

The population $P \in [0,1]$ quantifies the deviation between the prepared and ideal GHZ state ($\phi=0$) for the diagonals of the density matrix by measuring the state in the $Z$-basis, and computing the combined probabilities of measuring either the all-$0$ or all-$1$ states: 
 \begin{equation}
     P := \frac{1}{2}\left(p_{00...0}+p_{11...1}\right).
 \end{equation}

The Hellinger fidelity serves as a second figure of merit for the quality of the prepared state. Given two probability mass functions $p, q$ over $2^n$ states, the Hellinger fidelity is defined as 
\begin{equation} \label{eq:hellinger}
    H_F(p,q) := \left( \sum_{i=0}^{2^n-1} \sqrt{p_i q_i} \right)^2 \,.
\end{equation} 
When $q$ is the ideal GHZ state distribution, it is related to the population $P$ via ${H_F = P/2 + \sqrt{p_{00...0} \, p_{11...1}}}$. We remark that $H_F = P$ if and only if $p_{00...0}=p_{11...1}$.

\subsection{Experimental results}
We experimentally characterize the quality of the resulting GHZ state as a function of entangled-qubit number and other controllable parameters such as the number of flags, and inclusion/exclusion of error suppression in addition to the error-detection routine. For each experimental configuration, following the generation of the GHZ state over selected data qubits, we perform parity checks using the available flag qubits. All qubits (data and flag) are then measured in a characterization procedure (full measurement of the data qubits is not required when the generated state is to be used in a downstream circuit).

Measuring a flag in the $Z$-basis reveals the parity of the two data qubits; odd-parity shots result in a measurement outcome $-1$ and that shot is flagged as having a likely bit-flip or amplitude-damping error before being discarded. Each size of GHZ state preparation circuit is executed with the maximal number of type-0 flags, $f_\text{max}$, available for that size. In post-processing we post-select the samples according to $l$ flags up to the maximal number. While there are $\binom{f_\text{max}}{l}$ subsets of $l$ flags, we present the values corresponding to the subset achieving the maximal MQC fidelity. We additionally apply readout error mitigation~\cite{mundada2023experimental} during the post-processing step by first aggregating and readout-error-mitigating the parity-check flag bits, post-selecting, and then mitigating the remaining bits. We found this readout-error-mitigation scheme to yield optimal performance.

In Fig.~\ref{fig:layout_and_gen_circ}(c), we show the experimental MQC fidelity achieved for GHZ states of varying sizes as a function of the number of type-0 flags. In all cases, we find genuine multipartite entanglement above the critical $50\%$ threshold and observe that MQC fidelity grows with the number of flag qubits before exhibiting saturation.  We achieve maximum fidelities $>78\%$ for up to $54$Q, and $>55\%$ fidelity for $75$Q when using all available flag qubits for post-selection. Notably, the discarded fraction of shots, shown in Fig.~\ref{fig:layout_and_gen_circ}(d), reveals relatively modest error rates.  

 In Fig.~\ref{fig:layout_and_gen_circ}(e) and (f), we present the components of the MQC fidelity, coherence (see App.~\ref{app:ghz} for parity oscillations) and population, respectively. The results demonstrate the effectiveness of the flags in significantly boosting the fidelity and population, with the latter approaching unity even at $75$Q. We emphasize that there is a consistent trend across all system sizes---the fidelity increases with the number of flags being used in post-selection.  We observe that the quality increase of the coherence with the number of flags is low compared to that of the population, suggesting a considerably dephased state at the largest system size. This is in line with our expectation that the $Z$-basis parity checks detect $X$-type and $Y$-type errors, and not $Z$-type errors.

As a complementary measure, we present the Hellinger fidelity of the same GHZ sizes as a function of the number of flags used in post-selection in Fig.~\ref{fig:layout_and_gen_circ}(g). We observe the same trend of increasing fidelity with the number of flags, before saturation in the measured fidelity.  The Hellinger fidelities are close in value to the population, and approach unity, implying balanced $p_{00...0}$ and $p_{11...1}$, each approaching the ideal probability of $0.5$.  This result further confirms the efficacy of sparse parity checks against $X$-type and $Y$-type errors. 

We note that type-1 flags can also be added to the GHZ state preparation, as well as to the intermediate GHZ in supporting the long-range CNOT gate in the unitary entangle-disentangle protocol. In both cases, we show net negative benefits due to the added overhead of SWAP gates (see App.~\ref{app:ghz} and App.~\ref{app:cnot}, respectively).

Next, we disambiguate the role of error suppression and error detection in the experimental results presented above, giving insights into the relative strength of coherent and incoherent errors. For that purpose, we compare the performance of generating a 45Q GHZ state where the same sets of pre-transpiled circuits are executed back-to-back using either error detection and error suppression in combination or error detection alone. 
The MQC fidelity, Hellinger fidelity, and discarded fraction of shots as a function of number of flags used in post-selection on \texttt{ibm\_brisbane} are shown in Fig~\ref{fig:against_qiskit_45q}. Both MQC and Hellinger fidelities indicate that when exclusively using error detection (i.e., omitting the error suppression pipeline in circuit execution), GHZ state preparation is dramatically degraded. For example, even when all $5$ flags are used, the MQC fidelity achieved using error detection alone is well below the $0.5$ threshold. We remark that the device \texttt{ibm\_brisbane} used here for Fig.~\ref{fig:against_qiskit_45q} has a lower quality compared to the device \texttt{ibm\_fez} used for Fig.~\ref{fig:layout_and_gen_circ}.

The differences observed when including or excluding the error suppression pipeline indicate the presence of substantial coherent errors and readout errors which are effectively reduced through the pipeline, but are not captured through parity checks in the computational basis. Moreover, both fidelity measures demonstrate the efficacy of the error suppression pipeline at this scale, where the quality of the generated state saturates with 3 flags only, and the diagonals of the density matrix (captured by the Hellinger fidelity) approach their ideal values; high-frequency and stochastic $Z$ errors appear to dominate the residual infidelity. Examination of the discarded fraction in Fig~\ref{fig:against_qiskit_45q}(c) further shows little difference between the two trial cases.  This is consistent with the knowledge that error suppression is efficient at reducing temporally correlated error processes but inefficient at handling stochastic bit-flip and amplitude-damping errors, and that parity checks are insensitive to dephasing.  

\section{Discussion} \label{sec:discussion}

In this work, we have demonstrated that combining error suppression with QEC primitives on unencoded qubits can deliver substantial benefits in the computational capabilities of contemporary quantum computers.  Through a series of experiments on commercial superconducting processors, we demonstrate record-setting capabilities in both long-range CNOT gates and the generation of maximally entangled GHZ states. 

The protocols implemented are tailored to specific application modules (CNOT teleportation and GHZ state preparation), but leverage a common framework of low-overhead error detection in combination with overhead-free error suppression. Through our experiments, sparse parity checks prove a highly efficient QEC primitive that can be integrated into key algorithmic subroutines, despite the fact that circuit design incorporating error detection is not trivially generalized. Experiments also paint a consistent picture that the combination of the error-detection QEC primitive with error suppression ensures that both incoherent and coherent error processes can be efficiently suppressed.
  
Incorporating the use of QEC primitives delivers a new level of flexibility in the design of circuit modules which exhibit favorable properties for execution on current and near-term devices. For instance, compared to a previously demonstrated measurement-based protocol, the long-range CNOT-gate protocol we introduce trades $n-3$ measurements for the same number of CNOT gates, see Tab.~\ref{tab:overhead_teleported_cnot}. The measurement-based protocol is highly sensitive to single-shot readout errors, since all of the GHZ qubits are in superposition prior to measurement, and the measurement result will inform the feed-forward operations necessary to implement the correct quantum logic. Most contemporary quantum computers exhibit single-shot readout errors that are significantly higher than two-qubit-gate errors (by $3-6\times$); consequently, our new protocol is better suited for near-term systems. The trade-off comes with a circuit depth that scales linearly with the teleportation path length---this drawback is not expected to be significant for paths with $\lesssim 50$ qubits on current-generation devices, but can considerably limit the gate fidelity with longer teleportation paths (see App.~\ref{app:mixed_style} for scaling the protocol). Moreover, the overall measurement statistics of our protocol will be concentrated on a small number of bit-strings (independent of whether the disentangled qubits are measured), making the measurement distribution amenable to readout error mitigation and providing a further benefit in implementation. This contrasts the inherent ineffectiveness of readout error mitigation for the measurement-based protocol; see App.~\ref{app:bitstring_spreading}.

\begin{table}[t!]
\centering
\begin{tabular}{c|ccc}
\hline
 Protocol & Depth & $N_\text{CNOT}$ & $N_\text{msmt}$ \\
 \hline
 \hline
 Unitary entangle-distentangle & $n-1$ ($n$ even) & \multirow{2}{*}{$2n-2$} & \multirow{2}{*}{3} \\
 without error detection & $n$ ($n$ odd) & &\\
  \hline
  Unitary entangle-distentangle & $n-1$ ($n$ even) & \multirow{2}{*}{$2n-2$} & \multirow{2}{*}{$n$} \\
with error detection & $n$ ($n$ odd) & &\\
   \hline
    Measurement-based & $2$ & $n+1$ & $n$\\
   \hline
\end{tabular}
\caption{
\textbf{Overhead for long-range CNOT.} 
Here, $n$ is the size of the shortest-path, linear-chain GHZ state, as well as the distance (measured in qubits) between the control and target qubits of the long-range CNOT. Depth refers to the two-qubit-gate depth of the circuit. $N_\text{CNOT}$ and $N_\text{msmt}$ stand for the number of local CNOT gates and measurements, respectively. %
}
\label{tab:overhead_teleported_cnot}
\end{table}

 Our results make clear that the combination of deterministic error suppression and QEC primitives that stop short of full encoding can provide substantial computational enhancement with far lower overhead than full fault-tolerant QEC.  For instance, recent work published while completing this paper~\cite{reichardt2024atomcomputing} showed the use of logical encoding of an error-detection code in order to produce a 24-qubit error-detected GHZ state with approximately 90\% fidelity;  doing so required 59 physical qubits including 35 redundant qubits used in encoding, an overall (including ancillas) qubit overhead ratio of $(59-24)/24 \approx 146\%$ and 93\% discard rate. In our work, although not protected against incoherent phase errors, a comparable 27-qubit GHZ state also achieving 90\% fidelity requires an overhead ratio of only 7\% and just 20\% shots discarded.  At the maximum scale, our work uses 84 physical qubits to encode a 75-qubit error-detected GHZ state, requiring only 9 flag ancillas, amounting to a substantially lower overall qubit overhead ratio of $\approx 12\%$ with 78\% of shots discarded. Although both kinds of error detection with post-selection have fundamental limits concerning scalability, our adoption of QEC primitives on the physical level has a much lower overall overhead.
 
 These new results address the fact that most users of contemporary quantum computers will not yet see the net computational advantage when leveraging full logical encoding, despite the resounding success of many recent proof-of-principle experiments and the greater generality of logical encoding for arbitrary applications in the long term. Resource overhead in both QPU runtime and qubit count associated with QEC remain challenging and quickly overwhelm the physical device sizes available today. We therefore believe that our observations may help craft guidelines for the design of near-term algorithmic subroutines which deliver expanded computational capability over the best currently achievable routines by adopting the most impactful and efficient QEC primitives.  

Looking forward, we suggest two areas for future investigations to further improve the long-range CNOT gate. First, the generation of the linear-chain GHZ can be parallelized via the fusion of smaller GHZ states into a large one connecting the control and target qubits. 
Second, the disentangling of the GHZ into a distant Bell pair can also be parallelized with multiple branches of unitary disentangling, followed by projective disentangling of the corresponding root qubits. These mixed styles of unitary and projective preparation and disentangling of the linear-chain GHZ allow for a progression from the unitary entangle-disentangle limit ($\mathcal{O}(n)$-depth, $\mathcal{O}(1)$-measurements without error detection) to the measurement-based limit ($\mathcal{O}(1)$-depth, $\mathcal{O}(n)$-measurements): see App.~\ref{app:mixed_style}. Such a protocol may yield the optimal balance between two-qubit-gate and readout errors, while restricting the circuit depth, and thus optimizing both the fidelity and the execution speed of the constructed long-range controlled-unitary gate to be applied in e.g., QEC~\cite{LDPC_2024}. It also enables the protocol to scale to much longer gate teleportation paths. Finally, we emphasize that our novel unitary entangle-disentangle protocol can also serve as a useful QEC primitive itself for bi-layer architectures of solid-state platforms~\cite{LDPC_2024}.

\subsection*{Acknowledgments}
We are grateful to all other colleagues at Q-CTRL whose technical, product engineering, and design work has supported the results presented in this paper. We would also like to thank Alireza Seif, Elisa B\"{a}umer, and Vinay Tripathi for helpful discussions.

\section*{Supplementary Information}
\begin{appendix}

\section{Additional teleportation circuits}
\label{app:additional_circuits}
In this Appendix, we first provide a fully unitary disentangling strategy for the unitary entangle-disentangle protocol for long-range CNOT, providing more insight into how entanglement can be alternatively removed from the system with a similar overhead. Subsequently, we provide a general strategy to scale up the unitary entangle-disentangle protocol and to allow a more flexible trade-off between measurement and gate errors. Then, we extend our entangle-disentangle protocol for long-range CNOT gates to the long-range quantum fan-out gates and Toffoli gates. Lastly, we provide an alternative derivation of the measurement-based long-range CNOT protocol demonstrated by B\"{a}umer {\em et al.}~\cite{baumer2024efficient}.

\subsection{Fully unitary disentangling step for long-range CNOT}\label{app:fully_unitary_disent}

We recall from the unitary entangle-disentangle protocol described in Sec.~\ref{sec:cnot} and depicted in Fig.~\ref{fig:teleported_cnot} that the intermediate qubits in the GHZ state are disentangled unitarily following a circuit that mirrors the entangling one---the direction of the local CNOT gates are all towards the ends of the gate teleportation path, leaving the residual Bell pair on the ends untouched. The overall circuit thus resembles the MQC coherence circuit. A consequence of this symmetrical disentangling step is a ``root'' GHZ data qubit that remains in isolation, which should be projectively disentangled by an $X$-basis measurement. As will become clear below, this disentangling approach is important if one wishes to parallelize the disentangling step in the pursuit of a shallower circuit; see App.~\ref{app:mixed_style}.

\begin{figure}[!ht]
\centering
\includegraphics[width=0.95\columnwidth]{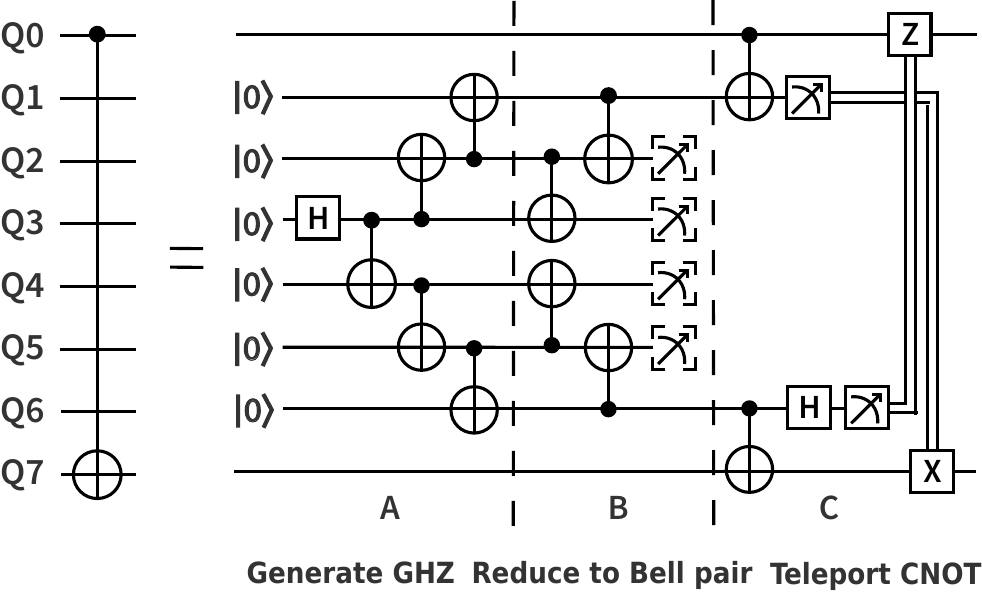}
\caption{
\textbf{Unitary entangle-disentangle protocol for long-range CNOT with a fully unitary disentangling step.} 
Here and throughout, single lines indicate qubits, double lines represent the flow of classical bits, and measurements with a dashed border are optional and required only for error detection. The protocol steps are separated by vertical dashed lines and indicated with the letters A, B, and C. The disentangling step B is fully unitary, in contrast to steps B-C in Fig.~\ref{fig:teleported_cnot} in the main text.
}
\label{fig:fully_unitary_disentangling}
\end{figure}

Alternatively, the reduction to a residual Bell pair can be achieved with a fully unitary circuit, by reversing both the ordering and direction of the local CNOT gates in the disentangling step compared to the same step in Fig.~\ref{fig:teleported_cnot}. This observation is depicted in Fig.~\ref{fig:fully_unitary_disentangling}. In particular, in step B, qubits towards the middle of the linear-chain GHZ are disentangled first, followed by the qubits towards the ends of the chain being disentangled. As we take advantage of the residual Bell pair as control qubits for the disentangling, the local CNOT gates for disentangling are all directed toward the center. Concretely, the state of the intermediate qubits after step B is 
\begin{equation}
    \frac{1}{\sqrt{2}}(\ket{00}+\ket{11})\otimes \ket{0}^{\otimes (n-2)} \,.
\end{equation}
All qubits, except the two at the ends, are in the ground state. Measuring these ground-state qubits allows for error detection. This alternative disentangling strategy is useful when we try to disentangle qubits connecting to the edges of the teleportation path; see Fig.~\ref{fig:quantum_fan_out} in App.~\ref{app:fan_out}. A consequence is that the $1$ measurement for projective disentangling is now traded for $1$ additional local CNOT gate during disentangling. This strategy thus shares very similar overhead with the one presented in Sec.~\ref{sec:cnot} (see Tab.~\ref{tab:overhead_teleported_cnot}), and it maintains the circuit depth. Therefore, we do not expect this alternative strategy to perform differently, especially for gate teleportation paths $\gtrsim 10$ qubits. However, a drawback is that it disallows a straightforward generalization to parallelize the disentangling step for a shallower circuit.

\begin{figure}[!]
\centering
\includegraphics[width=0.95\columnwidth]{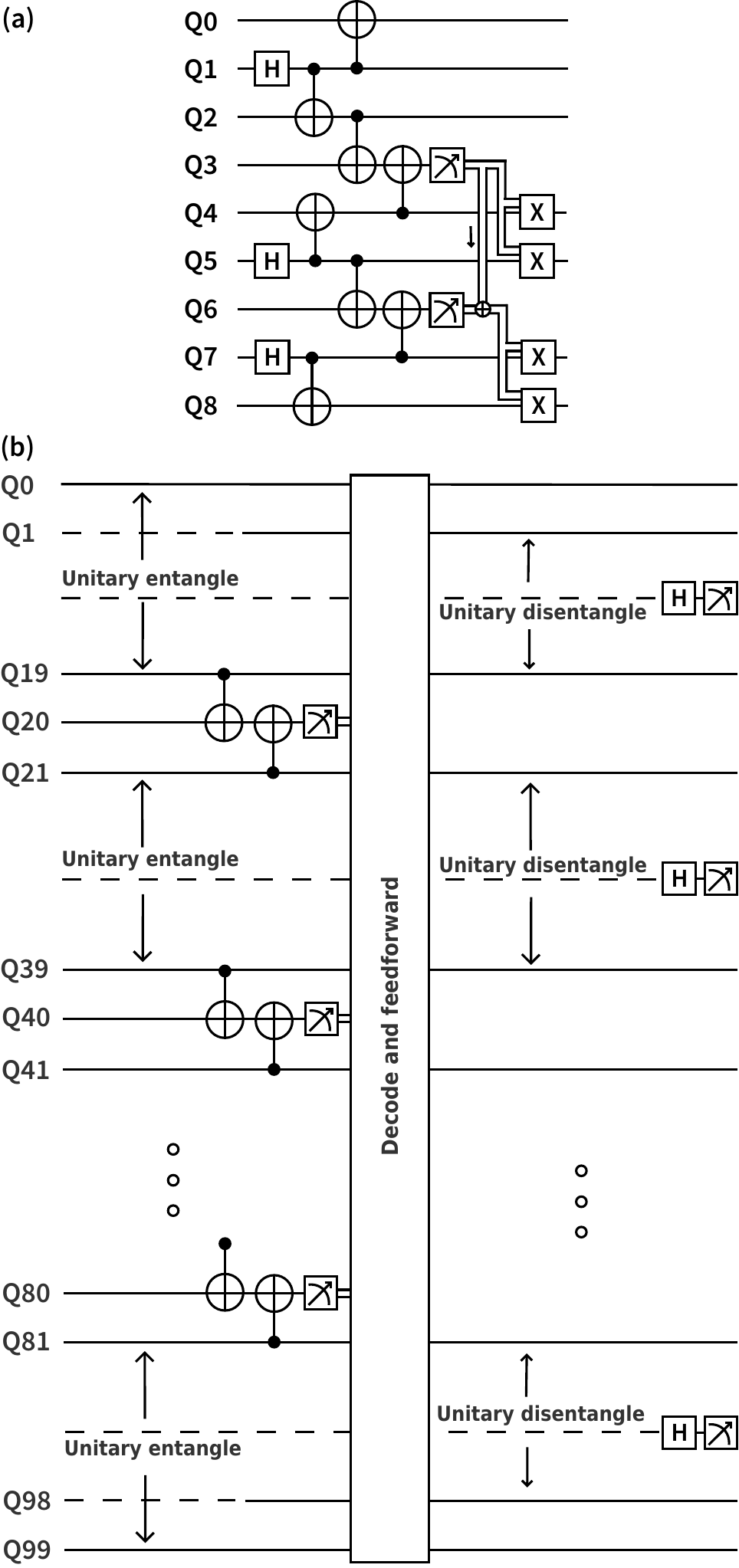}
\caption{
\textbf{Parallelization of the entangling and disentangling steps to scale up the long-range CNOT and to optimize for general error budgets and gate speed.} 
(a) An example of fusing a 3-qubit GHZ state (Q0--Q2) and 2 Bell pairs (Q4, Q5, and Q7, Q8) into a 7-qubit GHZ states. The small arrow indicates the direction of the flow of the classical bit. 
(b) Protocol to create a distant Bell pair over qubits Q0 and Q99. Select qubits are omitted for convenience (indicated by vertical double-arrowed lines and vertically oriented three open circles), as are certain operations (indicated by horizontal dashed lines). The middle box symbolizes the decoding operation that accepts all the mid-circuit measurement outcomes, and the feedforward operations necessary for the creation of the intermediate GHZ state on all other qubits. At the end of the circuit, the flow of the classical bits of the final measurement outcomes are omitted, since they are only relevant for the parity of the residual Bell pair on Q0 and Q99.}
\label{fig:mixed_style}
\end{figure}

\subsection{Designing circuits for general error budgets and scaling up} \label{app:mixed_style}

We continue our focus on the long-range CNOT gate and illustrate a general strategy to more flexibly balance measurement errors and gate errors, resulting in a family of circuits interpolating between the linear-depth, unitary entangle-disentangle protocol and the constant-depth, measurement-based protocol. The ability to flexibly design the circuit is also crucial for scaling up the unitary entangle-disentangle protocol for the long-range CNOT since its linear-depth circuit presented in Sec.~\ref{sec:cnot} will become prohibitively deep as the gate teleportation distance increases. It also allows one to optimize for the gate speed (circuit depth) critical for acting as a QEC primitive. The underlying principles discussed here can be applied to other non-local entangling gates as well. 

As discussed in Sec.~\ref{sec:discussion}, the disentangling step can be parallelized into multiple branches, with each root qubit projectively disentangled at the end of the step. If the parallelization is uniform, across a teleportation path length $n\approx k m$, the unitary disentangle step can be applied to $k$ groups of $\sim (m-1)$ qubits and $k-1$ ancillas. For instance, for a teleportation path across 100Q, the unitary disentangling step of the circuit can be applied to each of the contiguous subgroups of $\sim 19$ qubits: Q1--Q19, Q21--Q39, $\dots$, Q81--Q98, leaving behind a Bell pair on Q0 and Q99. Hence, the two-qubit-gate depth of the disentangling step is $10$, and $5\%$ of the qubits are projectively disentangled.

However, at large scales, the overall deep circuit (the entangling step remains an $(n/2)$-depth circuit) will wash out any derived benefits from parallelizing the disentangling step alone. Therefore, it is necessary to simultaneously parallelize the entangling step, so that the overall circuit depth is reasonably small (two-qubit-gate depth $\lesssim 50$ on current-generation devices). To that end, we can perform fusions of smaller GHZ states; see e.g., Ref.~\cite{de_Bone_2020}. One way to perform the fusion is to use ancillas to measure the parity of every neighboring small GHZ state, decode the ancilla measurements, and apply the corresponding $X$ gates on the qubits of the corresponding small GHZ states, as exemplified in Fig.~\ref{fig:mixed_style}(a) to fuse together a 7-qubit GHZ state on \{Q0, Q1, Q2, Q4, Q5, Q7, Q8\}. The decoding and feedforward operations (corrections) resemble that of the bit-flip repetition code with neighboring $ZZ$ syndrome measurements. In general, if the parallelization is uniform such that each of the $k$ branches in both the entangling and disentangling steps share about the same number of qubits, and across a teleportation path length $n\approx k m$, then the depth decreases from $\mathcal{O}(n)$ (precisely, $n-1$ if $n$ is even, and $n$ if $n$ is odd) to $\mathcal{O}(m)$ (approximately, $m$), while $\mathcal{O}(k)$ (precisely, $2k-1$) qubits would be measured in the absence of error detection. In the maximal parallelization (fusion) limit, we obtain the measurement-based preparation of a GHZ state~\cite{Andersen_2020, Moses_2023, chen2023nishimori, baumer2024efficient, hashim2024}.

Together with this fusion technique to parallelize the entangling step, we exemplify the creation of a Bell pair between Q0 and Q99 in Fig.~\ref{fig:mixed_style}(b): Parallelizing the entangling step, ancillas are placed between the small, unitarily prepared GHZ states to be fused together. For instance, Q20 is the ancilla between two small GHZ states on Q0$-$Q19 and Q21$-$Q39. Ancilla measurements are decoded, followed by $X$ gates feedforwarded onto the qubits of the corresponding small GHZ states. At this point, Q0$-$Q99 forms a large GHZ state. We then parallelize the disentangling step with simultaneous unitary disentangling, with each root qubit projectively disentangled. The final measurement outcomes collectively (by addition modulo 2) determine the parity of the residual Bell pair. The circuit has $\sim 10\%$ qubits measured, and a two-qubit-gate depth of $20$. As a comparison, the unitary entangle-disentangle limit to create the same distant Bell pair would have a currently prohibitive depth of $99$, and the measurement-based limit would have a depth of $2$.

In summary, a progression from linear-depth, unitary entangle-disentangle protocol, to a constant-depth, measurement-based protocol, while keeping the overall circuit reasonably shallow ($\lesssim 50$ depth in the near term), can be realized by parallelizing both the intermediate GHZ state generation and its reduction to a distant Bell pair. The extent of parallelization (namely, the size of smaller GHZ states to be fused, and the size of the subgroup of entangled qubits to be unitarily disentangled) determines the percentage of qubits being measured, allowing for a flexible balance between readout errors and gate errors. We finally remark that the advantage of having such a parallelization is absent in SWAP-based methods (e.g., Fig.~6.Ia in Ref.~\cite{baumer2024efficient}).

\subsection{Long-range quantum fan-out gates and Toffoli gates} \label{app:fan_out}

\begin{figure}[!hbt]
\centering
\includegraphics[width=1\columnwidth]{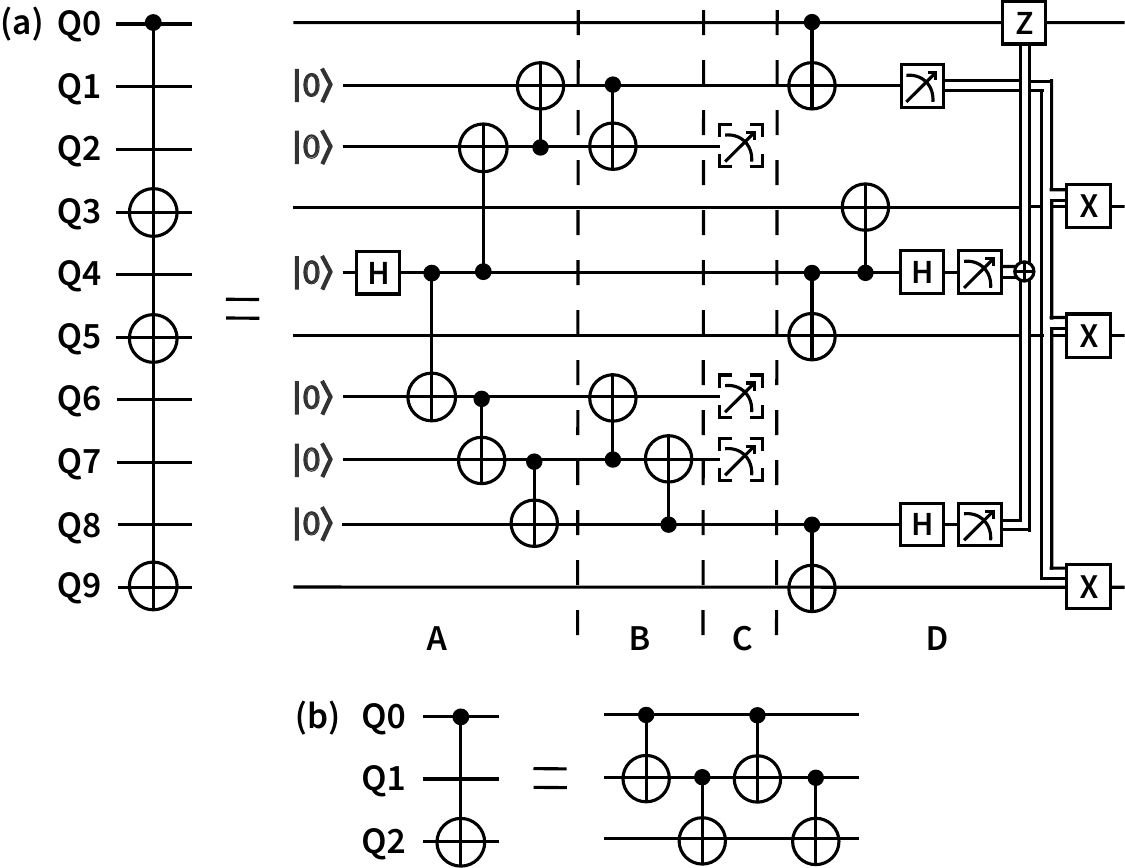}
\caption{
\textbf{Unitary entangle-disentangle protocol for long-range quantum fan-out gates}. 
(a) All qubits except the control (Q0) and the targets (Q3, Q5, and Q9) are initialized to $\ket{0}$. Measurements with a dashed border are optional and required only for error detection. The protocol steps are separated by vertical dashed lines and indicated with the letters A, B, and C.
(b) The non-local CNOTs appearing in (a) are implemented in a non-teleportation manner---for example, a CNOT gate between Q0 and Q2 skipping Q1 can be decomposed into a circuit with local CNOT gates on the right-hand side, regardless of the initial state of the skipped qubit.
}
\label{fig:quantum_fan_out}
\end{figure}

The three-step framework utilized in the App.~\ref{app:extended_derivation_dynamic_cnot} and in Sec.~\ref{sec:cnot} can be generalized to long-range multi-target controlled gates. We demonstrate this by providing a protocol for long-range (non-local) quantum fan-out gates, i.e., C$XX\dots X$ gates with one control and multiple non-local targets, which may serve as a competitive alternative to the measurement-based counterpart~\cite{baumer2024_more_gates, Paul2013}, especially on near-term devices, and further enable efficient error detection. Non-local quantum fan-out gates are useful for quantum memory access~\cite{allcock2023, Xu_2023} and the execution of Boolean functions~\cite{allcock2023, Hoyer2005}.

The protocol proceeds as follows. (1) Find the shortest path connecting all control and target qubits. Generate a GHZ state on this path, but skip all control and target qubits. Let $m$ denote the number of target qubits, and $n$ the size of the GHZ state. (2) Identify $k\leq m+1$ GHZ data qubits, each of which neighbors at least one control or target qubit. Every control and target qubit should have one such neighboring GHZ data qubit (multiple parties can share a common, neighboring GHZ data qubit, hence $k\leq m+1$; if no sharing is possible, $k=m+1$). Reduce the $n$-qubit GHZ state down to that $k$-qubit GHZ state. (3) Perform LOCC on the $m+1$ non-local parties based on the $k$-qubit GHZ state and achieve the long-range fan-out gate. In particular, the local operations consist of one local CNOT from the control to its GHZ data qubit neighbor, and $m+1$ native CNOTs from each target to their corresponding GHZ data qubit neighbor, followed by $1$ bit of classical information exchanged between the control and every target.

\begin{figure*}[!htbp]
\centering
\noindent\includegraphics[width=1\textwidth]{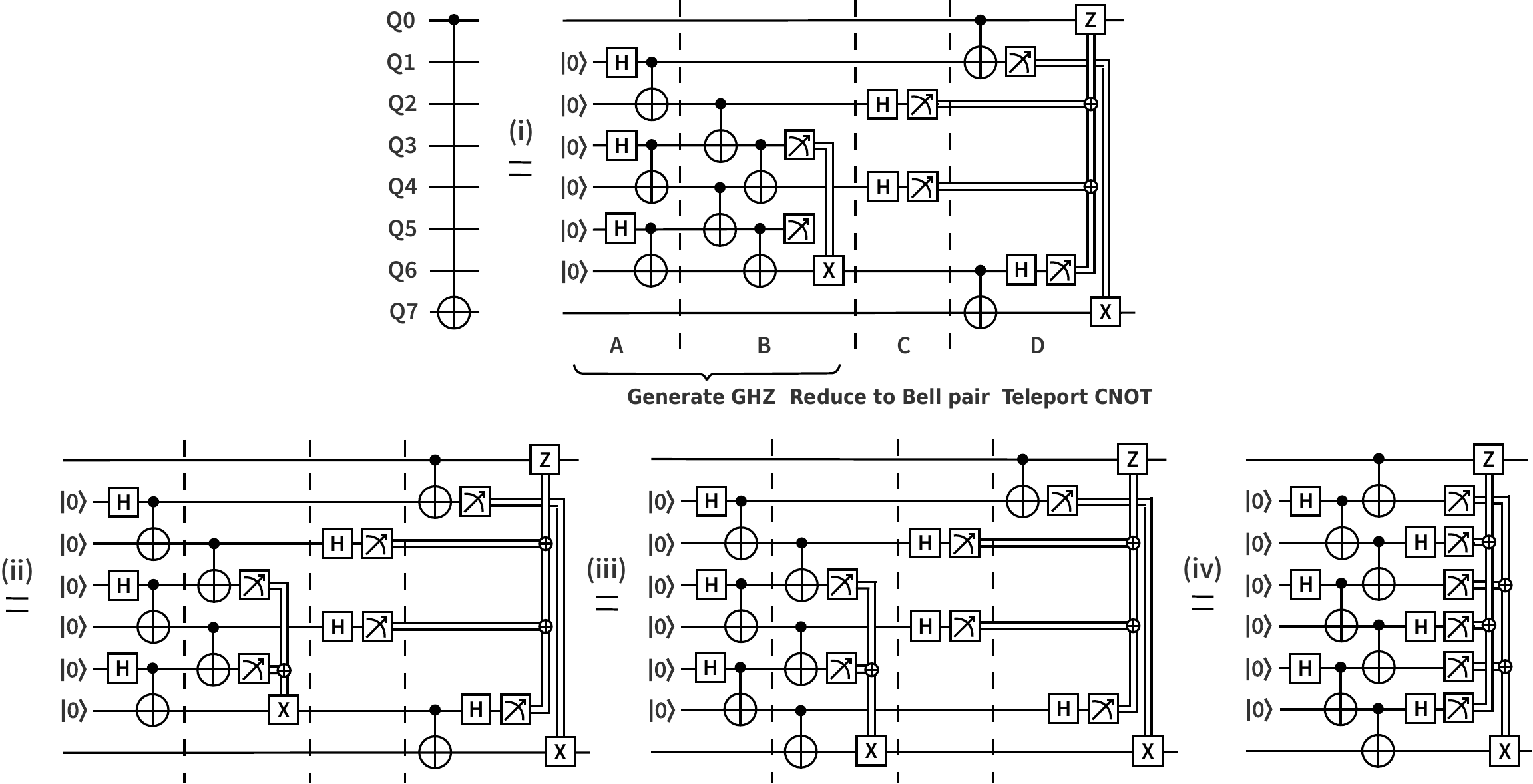}
\caption{\textbf{An alternative derivation of the measurement-based protocol~\cite{baumer2024efficient}}. 
(i) A long-range CNOT acting on Q$0$ (control) and Q$7$ (target), with Q1--Q6 all initialized to $\ket{0}$ separating the two. Vertical dashed lines (used throughout the derivation) and letters separate and denote different steps of the circuit explained in details in the Appendix. 
(ii) We invoke the principle of deferred measurements on the second layer of CNOTs in step B of the circuit, (iii) push the conditional-feedforward $X$ gate in step B through the second CNOT between Q6 and Q7, and (iv) merge all conditional-feedforward $X$ gates to recover the measurement-based protocol.
}
\label{fig:dynamic_cnot_appendix}
\end{figure*}

In Fig.~\ref{fig:quantum_fan_out}, we depict the unitary entangle-disentangle protocol for a long-range quantum fan-out gate on $3$ non-local target qubits, which can be readily generalized to an arbitrary number of targets and arbitrary range.  In step A, all qubits except the control (Q0) and the targets (Q3, Q5, and Q9) form a GHZ state by unitary entangling. Subsequently in step B, we leave one GHZ data qubit neighboring each party untouched, namely, leave Q1 (neighboring Q0), Q4 (neighboring Q3 and Q5), and Q8 (neighboring Q9) untouched, and unitarily disentangle the rest of the GHZ data qubits, namely Q2, Q6 and Q7. In step C, the unitarily disentangled qubits in principle end in $\ket{0}$, and can be optionally measured (the dashed measurement symbols) for error detection. Lastly in step D, local operations on all parties based on the shared GHZ state, and classical communication between the control and each target are performed, achieving the long-range fan-out gate.

We remark that the generation of the GHZ state, and the reduction to a smaller GHZ state shared by each party are achieved completely unitarily. The unitary entangle-disentangle protocol retains a characteristic $\mathcal{O}(n)$-depth, $\mathcal{O}(1)$-measurement overhead, in contrast to the characteristic $\mathcal{O}(1)$-depth, $\mathcal{O}(n)$-measurement overhead of the corresponding measurement-based protocol~\cite{baumer2024_more_gates}.

Beyond long-range multi-target gates, we suggest that both the measurement-based protocol and unitary entangle-disentangle protocol can be adapted to achieve long-range multi-control (or most generally, multi-party) entangling gates. Taking the non-local Toffoli gate as an example, we first note a theorem presented by Eisert {\em et al.}~\cite{Eisert_2000} states that: $2$ shared Bell pairs and a total of $4$ bits of classical communication are necessary and sufficient for the local implementation of a non-local three-party quantum Toffoli gate (see Fig.~4 in Ref.~\cite{Eisert_2000}). A practical non-local implementation thereby relies on generating $2$ Bell pairs from $2$ GHZ states, before performing LOCC on the $3$ parties, although it still requires a local Toffoli gate. 

Specifically, we consider two scenarios of applying a non-local Toffoli gate. In the first scenario, the 2 control qubits are situated on two sides of (and do not neighbor) the target qubit. The gate can be readily realized by directly, unitarily generating 2 linear-chain GHZ states connecting each control to the target, before a corresponding unitary disentangling down to 2 Bell pairs, with each pair neighboring a control and the target.

In the second scenario, the 2 control qubits are situated on the same side of the target qubit, such that their shortest paths to the target overlap to some degree. For instance, on a linear chain of 10 qubits, Q0 and Q1 are the control qubits, while Q9 is the target of the Toffoli gate. This is challenging since we are required to obtain 2 Bell pairs, yet there appears to be no space for 2 GHZ states. Despite this complication, one can still achieve the non-local Toffoli gate with a unitary entangle-disentangle protocol by a linear-depth circuit. The trick is to apply local CNOT gates with skip connections as depicted in Fig.~\ref{app:fan_out}(b) (cascading it allows to skip more than 2 qubits if needed): The first GHZ state lies on $\{$Q2, Q4, Q6, Q8$\}$ serving as the entanglement resource for the control qubit Q0, while the second GHZ state lies on $\{$Q3, Q5, Q7$\}$ serving for the other control qubit Q1. The corresponding unitary disentangling can also be realized with local CNOT gates with the same skip connections, mirroring the entangling circuit. In general, the circuit depth of this unitary entangle-disentangle protocol is no more than approximately $4n$.

\subsection{Alternative derivation of the measurement-based protocol for long-range CNOT} \label{app:extended_derivation_dynamic_cnot}

We observe that the measurement-based protocol for the long-range CNOT~\cite{baumer2024efficient} also follows the three-step general framework of teleporting CNOT gates described in Sec.~\ref{sec:cnot} and depicted in Fig.~\ref{fig:teleported_cnot}(a). The protocol can be understood as fusing a linear chain of local Bell pairs (and an extra qubit placed in the $\ket{+}$ state if needed) into a GHZ state, reducing the GHZ state to produce a non-local Bell pair on the two endpoints of the path, followed by a teleported CNOT via LOCC based on the Bell pair shared by the control and target qubits.

More concretely, in Fig.~\ref{fig:dynamic_cnot_appendix}, we provide an alternate derivation of the measurement-based protocol of long-range CNOT demonstrated in B\"{a}umer {\em et al.}~\cite{baumer2024efficient} that makes the three steps in the general framework explicit. Our alternative derivation starts with step A, where Q1--Q6 form three Bell pairs. Subsequently in step B, entangling gates and measurements (all but the last one have feedforwards) fuse the Bell pairs into an intermediate GHZ state on Q1, Q2, Q4, and Q6. Later in step C, all but two GHZ data qubits (Q2 and Q4) are disentangled through $X$-basis measurements, producing a Bell pair between Q$1$ and Q$6$. Notably, the parity of the measured bit-string on the projectively disentangled qubits dictates the parity of the remaining Bell pair, thereby these measurement outcomes contribute to the conditional-feedforward $Z$ gate at the end of the circuit by addition modulo $2$. Finally, in step D, the teleported CNOT gate is achieved by LOCC based on the Bell pair. This alternative derivation can be generalized to any teleportation path length---for odd path length, we fuse an extra qubit in $\ket{+}$ state into the intermediate GHZ state formed in step B. We remark that in particular, the generation of the intermediate GHZ state is achieved partially through (steps A and B), and the reduction to Bell pair completely through (step C), mid-circuit measurement and feedforward operations. 

Lastly, we suggest that near-term devices are not likely to allow error detection with net gain using the measurement-based protocol, for the following reason: The minimal weight of the stabilizers of the intermediate state, immediately before the collective $Z$-basis measurements in the right-hand side of Fig.~\ref{fig:dynamic_cnot_appendix}, is $3$. For instance, $X_1 Z_2 X_3$ is one such stabilizer. Since the subsequent measurements of the circuit are all in the $Z$ basis, it requires the addition of an ancilla qubit to measure any of the stabilizers. Moreover, checking any of them requires significant gate overhead, especially for devices with limited connectivity. 

As a final remark, we note that the constant-depth, measurement-based protocol for long-range CNOT can be traced back to earlier works by Pham and Svore~\cite{Paul2013}, and Delfosse {\em et al.}~\cite{delfosse2021}.

\begin{figure}[!ht]
\centering
\includegraphics[width=0.73\columnwidth]{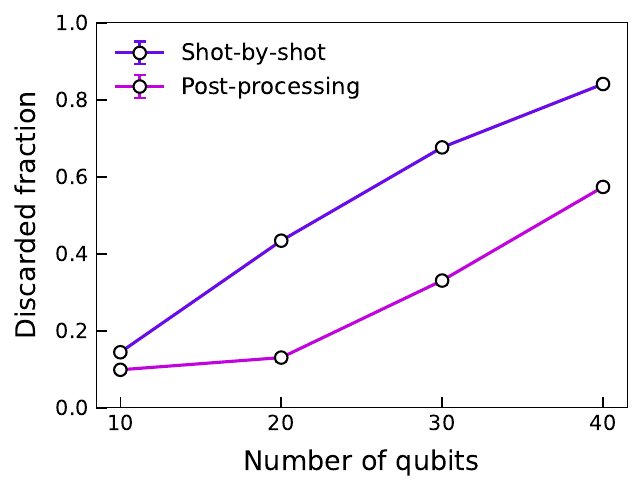}
\caption{\textbf{Discarded fraction of shots of unitary entangle-disentangle protocol for long-range CNOT when error detection is enabled.} The discarded fraction of shots corresponding to the results presented in Fig.~\ref{fig:teleported_cnot_results}.  Error bars show $\pm 1$ standard deviation of $50$ bootstraps.}
\label{fig:cnot_teleportation_discarded_fraction}
\end{figure}

\begin{figure*}[!ht]
\centering
\includegraphics[width=0.95\textwidth]{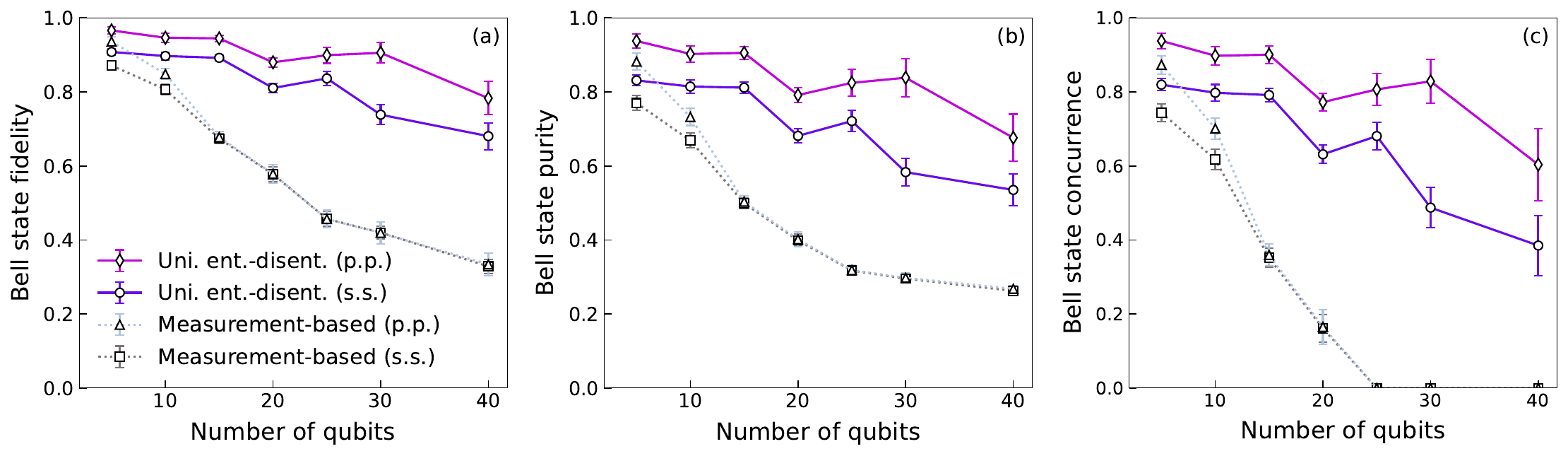}
\caption{
\textbf{Using the unitary entangle-disentangle long-range CNOT protocol to generate a distant Bell pair.} 
Two variants of the long-range CNOT protocol are used to generate a distant Bell pair---the unitary entangle-disentangle protocol (uni. ent.-disent.) and the measurement-based protocol. In both cases, the resulting Bell pair density matrix is reconstructed via state tomography. Performance metrics include (a) state fidelity, (b) purity, and (c) concurrence of the Bell pair, evaluated in both shot-by-shot experiments (s.s.) and post-processed data (p.p.). Error bars show $\pm 1$ standard deviation of $50$ bootstraps.}
\label{fig:cnot_teleportation_bell_pair}
\end{figure*}

\begin{figure*}[!ht]
\centering
\includegraphics[width=1\textwidth]{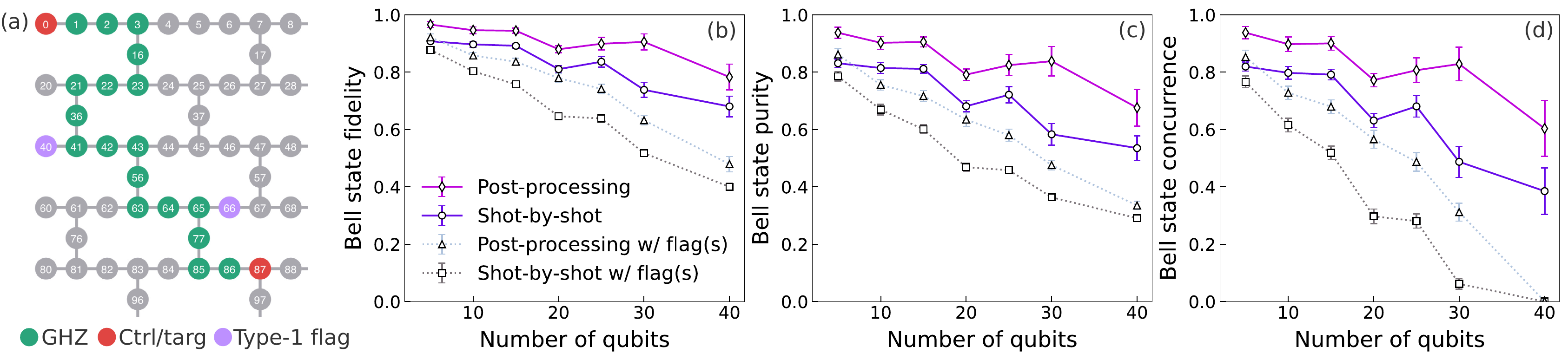}
\caption{
\textbf{Adding type-1 parity check flags to the unitary entangle-disentangle long-range CNOT protocol.} 
(a) An example of the long-range CNOT between the control and target qubits (red) across a linear chain of qubits forming an intermediary GHZ state (green). Type-1 parity-check flags (purple) are inserted to detect errors in the intermediary GHZ state. For example, the parity between Q41 and Q42 is checked by the type-1 flag qubit Q40. 
(b–d) The long-range CNOT protocol generates a distant Bell pair, benchmarked in terms of (b) fidelity, (c) purity, and (d) concurrence of the reconstructed Bell pair density matrix using post-processed data. 
Error bars show $\pm 1$ standard deviation of $50$ bootstraps.}
\label{fig:cnot_teleportation_typeIIflags}
\end{figure*}

\section{Long-range CNOT with error detection}
\label{app:cnot}

We provide additional information and experimental results for the novel long-range CNOT protocol introduced in Sec.~\ref{sec:cnot} in the main text. There, the average gate fidelity was used as the primary figure of merit to judge the efficacy of the experimentally realized operation. We therefore begin with a brief review of the gate fidelity estimation procedure (see also App.~C of Ref.~\cite{baumer2024efficient}).

\subsection{Average gate fidelity}
The average output fidelity, defined as the fidelity between the output states produced by two quantum channels
$\Lambda$ and $\Tilde{\Lambda}$ on a $d$-dimensional space averaged Haar-uniformly over all pure input states, can be shown to relate to the process fidelity as~\cite{Horodecki1999}
\begin{align}
    F_\text{gate}(\Lambda,\Tilde{\Lambda})=\frac{d \, F_\text{process}(\Lambda,\Tilde{\Lambda})+1}{d+1}.
\end{align}
As such, we estimate the process fidelity using Monte Carlo process certification~\cite{da_Silva_2011, baumer2024efficient}. According to the Choi-Jamiolkowski isomorphism, every quantum
operation $\Lambda$ on a $d$-dimensional space can be mapped to a density operator $\rho_\Lambda=(I\otimes \Lambda)\ketbra{\phi}$ with $\ket{\phi}=1/\sqrt{d}\sum_{i=0}^{d-1} \ket{i}\otimes \ket{i}$, called the Choi state. The process fidelity between the two channels $\Lambda$ and $\Tilde{\Lambda}$ is equal to the state fidelity between the two corresponding Choi states,
\begin{align}
    \label{eq:process_fid}
    F_\text{process}(\Lambda,\Tilde{\Lambda})=F(\rho_\Lambda, \rho_{\Tilde{\Lambda}}).
\end{align}
Furthermore, the state fidelity between two states $\rho$, $\sigma$, can be written as
\begin{align}\label{eq:fid_in_pauli_basis}
    F(\rho, \sigma) = \Tr(\rho\sigma) = \sum_i\frac{\rho_i\sigma_i}{d}, 
\end{align}
when at least one of the two states is pure. Here, ${\rho_i \equiv \Tr(\rho P_i)}$ and ${\sigma_i \equiv \Tr(\sigma P_i)}$, and $\{P_i\}$ are the set of $d$-dimensional Pauli operators.

As described in Ref.~\cite{da_Silva_2011}, a practical implementation of the quantum Monte Carlo state certification consists of preparing the complex conjugate of random product of eigenstates of
Pauli operators $P_i$, applying the
transformation $\Tilde{\Lambda}$ to the system, and finally measuring a random Pauli operator $P_j$ on each qubit, since the following holds true for the projection of the Choi states onto the Pauli basis~\cite{Nielsen_2002}:
\begin{equation}
    \Tr[(P_i\otimes P_j)(I\otimes \Tilde{\Lambda})\ketbra{\phi}] = \frac{1}{d}\Tr[P_i \Tilde{\Lambda}(P_j^*)].
\end{equation}
Here, $P_i$ and $P_j$ are two-qubit Pauli operators.

For an ideal CNOT gate, it is not hard to verify that only $16$ combinations of input Pauli eigenstate and measured Pauli operators have non-vanishing expectation values, as was also noted in Ref.~\cite{baumer2024efficient}. Therefore, when calculating the process fidelity of a noisy long-range CNOT gate to the ideal CNOT gate, only $16$ non-vanishing terms exist in Eq.~\eqref{eq:process_fid} upon expanding in the Pauli basis by Eq.~\eqref{eq:fid_in_pauli_basis}, and thereby up to $16$ circuits are needed for the gate fidelity estimation.

\subsection{Additional experimental results}

In evaluating the quality of the long-range CNOT, it is also essential to examine the fraction of discarded shots. In Fig.~\ref{fig:cnot_teleportation_discarded_fraction}, we present the discarded fraction of shots as a function of the number of qubits (or path length for the long-range CNOT). Shots are discarded if any of the $n-3$ intermediate qubits (see Eq.~\eqref{eq:intermediate_state_for_CNOT}) that should ideally be in the ground state are instead measured to yield $\ket{1}$. We present results for discarding shots on a shot-by-shot basis as well as after applying readout error mitigation to a statistical sample of shots in a post-processing step. The discard fraction is significantly reduced in the latter case, illustrating that, if available, statistical methods can significantly improve the yield of an error-detected gate.

In the main text, we use the average gate fidelity as a primary figure of merit to assess the performance of the unitary entangle-disentangle protocol. As a second figure of merit, here we evaluate the long-range CNOT operation by creating a non-local Bell pair and measuring the fidelity of the experimentally realized state. In more detail, we first initialize the control qubit in the $\ket{+}$ state, apply the long-range CNOT to create a non-local Bell pair, and then reconstruct the density matrix of the Bell pair using standard state tomography. The reconstructed density matrix can be obtained through 9 basis measurements of the state, and is manifestly Hermitian and unit-trace. Positive semi-definiteness is enforced by truncating its negative eigenvalues and rescaling its positive eigenvalues~\cite{Smolin2012}.
As shown in Fig.~\ref{fig:cnot_teleportation_bell_pair}, the unitary entangle-disentangle protocol consistently and significantly outperforms the measurement-based protocol in terms of the created Bell pair's fidelity, purity, and concurrence across the range of teleportation distances considered (up to 40 qubits).

Lastly, we consider whether additional parity checks with larger overhead can be used to further improve performance.
Ideally, the ancilla qubit involved in a parity check will be positioned adjacent to the pair of data qubits whose parity is to be checked. However, in practice, many circuits and device topologies do not allow for such minimal overhead parity checks, and the ancilla will be necessarily placed further away on the device. It is then interesting to investigate generalizations of such ancilla qubits for parity checks beyond minimal overhead. Due to the invariance of the GHZ state under qubit permutations, in principle only one additional SWAP gate is required for a parity check should the ancilla sit next to just one of the connected data qubits. We call such a parity check type-1, to distinguish them from minimal-overhead parity checks which we call type-0.

Due to the circuit topology over a linear chain of physical qubits, the long-range CNOT protocol does not support type-0 flags in general. In Fig.~\ref{fig:cnot_teleportation_typeIIflags} we therefore compare the performance of the protocol both with and without type-1 flags. We find that the type-1 flags lead to worse performance. The improvement due to the additional error detection and post-selection afforded by these parity checks is overwhelmed by the cost incurred by the additional circuit operations required to implement the type-1 parity checks.

\begin{figure*}[!ht]
\centering
\includegraphics[width=1\textwidth]{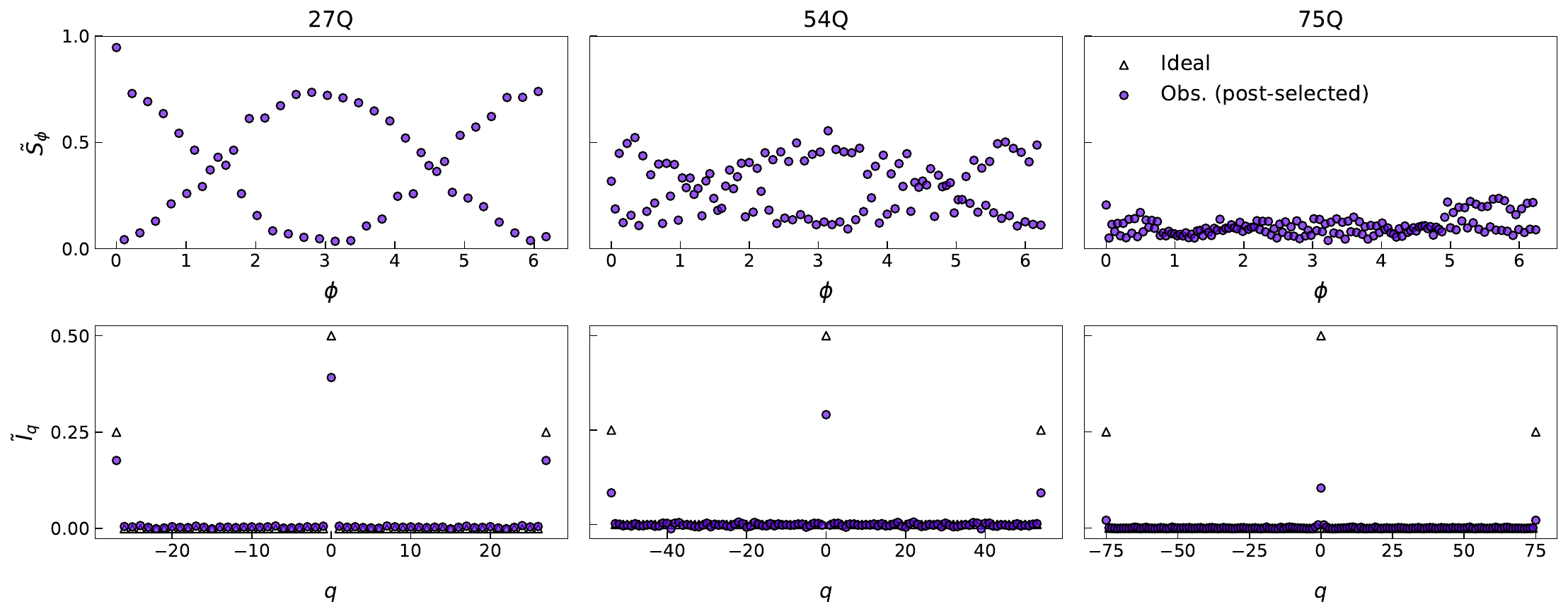}
\caption{\textbf{MQC parity oscillations.} (Top) The parity oscillations $\Tilde{S}_\phi$ as a function of the injected relative phase $\phi$ per GHZ data qubit, corresponding to the coherence measured in Fig.~\ref{fig:layout_and_gen_circ}(e) for three systems sizes, executed on \texttt{ibm\_fez}. 
The sampling of the phase $\phi$ is at the Nyquist limit for detecting the $\cos(n\phi)$-mode of the oscillation. (Bottom) The corresponding real discrete Fourier modes $\Tilde{I}_q$.
}
\label{fig:parity_oscillations}
\end{figure*}

\begin{figure}[!ht]
\centering
\includegraphics[width=0.8\columnwidth]{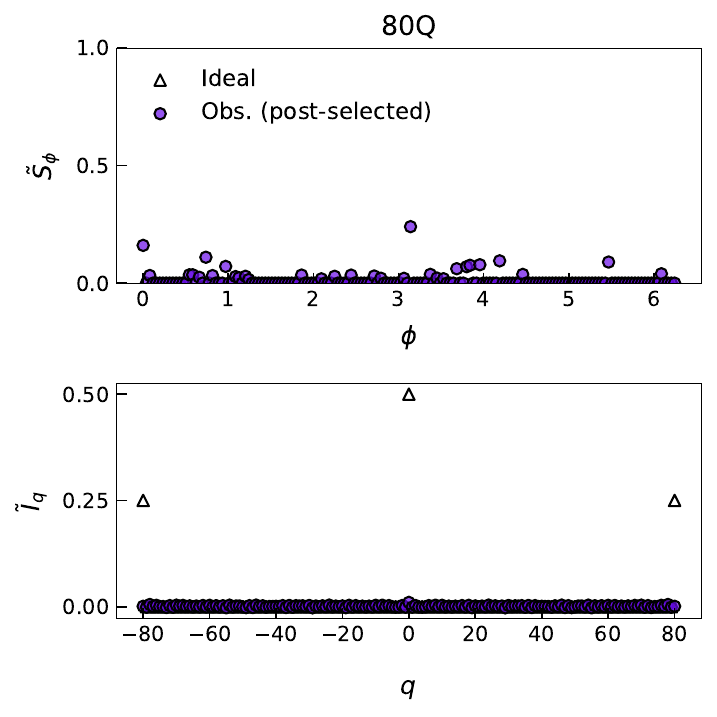}
\caption{\textbf{MQC parity oscillations for 80Q GHZ state with 10 type-0 parity check flag qubits.} (Top) The parity oscillation $\Tilde{S}_\phi$ as a function of the injected relative phases $\phi$ per GHZ data qubit, executed on \texttt{ibm\_fez}. (Bottom) The corresponding real discrete Fourier modes $\Tilde{I}_q$.}
\label{fig:parity_oscillations_80q}
\end{figure}

\begin{figure*}[!ht]
\centering
\includegraphics[width=0.93\textwidth]{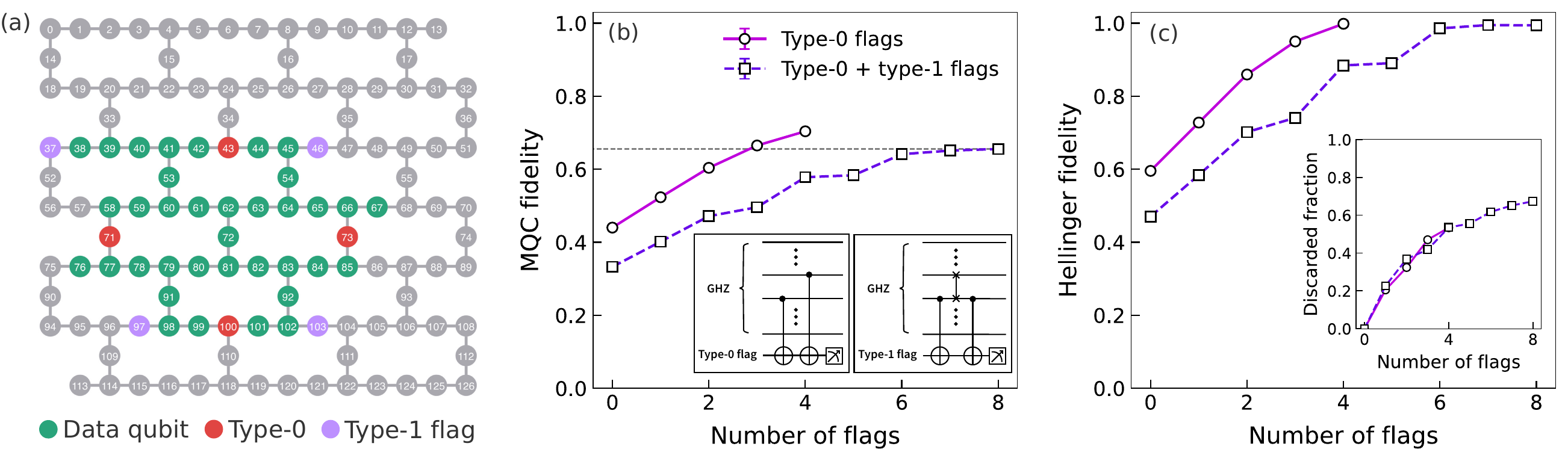}
\caption{
\textbf{GHZ state preparation with additional type-1 parity-check flags that require SWAPs.
}
(a) Layout of a 36-qubit GHZ state generated on \texttt{ibm\_brisbane}. Minimal-overhead type-0 flags, which require no SWAP operations, are shown in red, while type-1 flags, requiring a single SWAP each, are shown in purple. For example, the type-1 flag qubit Q46 uses a single SWAP operation to verify the parity of data qubits Q45 and Q54.
(b) The MQC fidelity of the 36-qubit GHZ state as a function of the total number of flags (including both type-0 and type-1). The dashed horizontal line represents the maximum fidelity achieved with all type-1 flags included, and the inset illustrates circuits for type-0 and type-1 parity-check operations.
(c) The Hellinger fidelity of the same 36-qubit GHZ state, also as a function of the total number of flags. The inset shows the fraction of discarded shots after post-selection, averaged across circuits. 
In both (b, c), the magenta curve with circle markers corresponds to the fidelity when only type-0 flags are used. Error bars represent $\pm 1$ standard error, though most are too small to be visible.
}
\label{fig:typeII_flags}
\end{figure*}

\section{GHZ state preparation with error detection}
\label{app:ghz}
In this Appendix, we provide additional experimental results for the GHZ state preparation experiments discussed in Sec.~\ref{sec:ghz} of the main text. 

We first provide further details on the state-preparation circuit. To prepare the GHZ states, the entanglement grows in a breadth-first-search manner: At each layer of the circuit, the set of already entangled qubits expands to include the set of qubits adjacent to them by applying local CNOT gates, resulting in a sub-linear-depth circuit.

We then briefly review the MQC formalism. Our aim is to measure the fidelity of the experimentally prepared and noisy GHZ state, $\tilde{\rho}$. By exploiting the simple structure of the ideal GHZ state, $\rho$, and making an assumption about the form of the noisy state, the MQC formalism allows the fidelity to be estimated using only $2n + 3$ experiments.

First, using the fact that the non-zero entries of the ideal GHZ state $\rho$ are at the corners in the computational basis, the fidelity of the prepared state is:
\begin{equation}
    F(\rho, \tilde{\rho}) = \frac{1}{2}(\tilde{\rho}_{0...0, 0...0} + \tilde{\rho}_{1...1,1...1} + \tilde{\rho}_{0...0,1...1} + \tilde{\rho}_{1...1,0...0})\,.
\end{equation}
The first two terms may be straightforwardly obtained using a single experiment; the population $P$, defined as ${P = \rho_{0...0, 0...0} + \rho_{1...1,1...1}}$, simply corresponds to the probability of measuring either the all-0 or the all-1 bit-strings after preparing the GHZ state. The second terms are collected into a single quantity, the coherence, $C$. Within the MQC formalism, the coherence may be obtained from the $n$-th real Fourier mode of the parity oscillation (interference pattern) using the relation $C = 2 \sqrt{I_n}$. Operationally, this is obtained by first creating a spectrum of phase-rotated GHZ states, then applying the inverse, non-phase-rotated GHZ preparation circuit, and finally measuring the all-$Z$ expectation values (see Fig.~\ref{fig:layout_and_gen_circ}(b)), which we Fourier transform.

In more detail, the phase-rotated state is 
\begin{equation}
    \rho_{\phi} = e^{-i M_Z \frac{\phi}{2}} \, \rho \, e^{i M_Z \frac{\phi}{2}} \,,
\end{equation}
where $M_z = \sum_{j=0}^{n-1} Z_{j}$ is the total magnetization operator. In the ideal case, the parity oscillation follows: 
\begin{align}
    \label{eq:GHZ_overlap_ideal}
    S_{\phi} &=
    \Tr( \rho_{\phi} \, \rho) 
    = \frac{1}{2}\left[1+\cos(n\phi)\right].
\end{align} 
For a general, noisy, prepared state, the parity oscillation is assumed to be band-limited in the MQC formalism, and admits the Fourier expansion
\begin{equation}
    \Tilde{S}_{\phi} =
    \Tr( \tilde{\rho}_{\phi} \, \tilde{\rho}) = \sum_{q=-n}^n e^{-i q \phi} \Tilde{I}_q \,.
\end{equation}
The maximal-frequency Fourier mode $I_n$ may be extracted by sampling the phase rotation angle at the Nyquist rate, ${\phi_j = \pi j / (n+1)}$ for ${j=0, ..., 2n + 1}$, and applying the inverse discrete Fourier transform:
\begin{equation}
    \label{eq:fourier_modes}
    \Tilde{I}_q = \frac{1}{2n+2}\Re \left( \sum_{j} e^{i q \phi_j} \, \Tilde{S}_{\phi} \right)\,.
\end{equation}
Importantly, although the coherence depends on only a single Fourier mode, resolving this requires $2n + 2$ experiments. Note that in the ideal case, $I_0 = 1/2$, and $I_n = I_{-n} = 1/4$, with all other $I_q = 0$ (cf., Eq.~\ref{eq:GHZ_overlap_ideal}). 

Next, in Fig.~\ref{fig:parity_oscillations}, we present experimentally-measured values for the parity oscillation $\Tilde{S}_{\phi}$ and the extracted Fourier modes $\Tilde{I}_q$ for $n=27$, $54$, and $75$, corresponding to the experiments presented earlier in Fig.~\ref{fig:layout_and_gen_circ}. We remark that at large system sizes, the prepared GHZ states are considerably dephased. In Fig.~\ref{fig:parity_oscillations_80q} we additionally present the parity oscillation for a prepared $80$Q GHZ state with $10$ type-0 flag qubits. The interference pattern is nearly entirely damped, and thus we do not claim to have prepared a state exhibiting genuine multipartite entanglement, despite its MQC fidelity at $F=0.511(2)$ with contributions from $C=0.064(3)$ and $P=0.958(2)$. We also note that the discarded fraction here with all $10$ flags is around $98\%$.  

Finally, as in App.~\ref{app:cnot}, here we also consider whether parity checks with non-zero overhead can be used to further improve performance. In particular, we consider the addition of type-1 parity checks requiring a single SWAP operation, together with the previously considered minimal-overhead type-0 flags, see Fig.~\ref{fig:typeII_flags}. As before, we find that the introduction of type-1 flags is not advantageous, since the decrease in fidelity due to the additional SWAP outweighs the improvement due to measuring an additional symmetry for error detection. We expect that the type-0 flags alone can already detect the majority of propagated errors, hence the marginal gain achieved by inserting additional type-1 flags is insufficient to overcome the cost of the additional SWAP.

\section{Insufficient bit-string counts ineffectuate readout error mitigation} \label{app:bitstring_spreading}

Readout error-mitigation techniques that rely on statistical inference~\cite{mundada2023experimental, mthree} are generally ineffective 
whenever the probability of the most probable bit-string(s) (in the measurement basis), $P_{\text{max}}$, satisfies $P_{\text{max}} \ll 1/N_{\text{shots}}$.
For instance, this is satisfied if the underlying (ideal) $n$-qubit state before simultaneous measurements of sufficiently many qubits is a uniform superposition across the exponential number of computational basis states, but is only sampled with a polynomial (in $n$) number of shots. Relevant to our work, three algorithms (circuits) fall into this category and thereby exhibit the phenomenon:
\begin{enumerate}
    \item The measurement-based preparation of $n$-qubit GHZ states~\cite{Andersen_2020, Moses_2023, chen2023nishimori, baumer2024efficient, hashim2024}. 
    The superposition state before simultaneous measurements of $\mathcal{O}(n)$ qubits is uniform, since each measurement on an ancilla has equal probabilities to project the corresponding pair of data qubits into the $\pm1$-eigenspace of $ZZ$. For that reason, we choose to focus on unitary state preparation.
    \item Fidelity estimation of an $n$-qubit GHZ state by sampling its stabilizers~\cite{flammia2011direct, Cao2023, chen2023nishimori, baumer2024efficient} discussed in Sec.~\ref{sec:2d_ghz_results}, especially with the full-weight, non-$Z$-type stabilizers which constitute half of all the $2^n$ stabilizers. We remark that any $X$-basis or $Y$-basis measurement on a GHZ data qubit yields $0$ or $1$ with equal probabilities, hence the uniform measurement statistics. For that reason, we choose the MQC fidelity as our primary fidelity measure.
    \item The measurement-based protocol of the long-range CNOT~\cite{baumer2024efficient} discussed and alternatively derived in App.~\ref{app:extended_derivation_dynamic_cnot}. We remark that any measurement in the circuit to generate or to disentangle the intermediate GHZ state, i.e., in steps B and C in Fig.~\ref{fig:dynamic_cnot_appendix}, results in an equal probability of obtaining $0$ or $1$ for each qubit, hence the uniform measurement statistics. Consequently, we observed across Fig.~\ref{fig:teleported_cnot_results}(b) and (c) that the measurement-based protocol does not benefit from additional readout error mitigation applied in post-processing for teleporting across $>10$Q.
\end{enumerate}

\end{appendix}

\bibliography{refs}

\end{document}

%% file: preamble.tex
\usepackage{graphicx}
\usepackage{dcolumn}
\usepackage{bm}
\usepackage[utf8]{inputenc} 
\usepackage[T1]{fontenc}    
\usepackage{hyperref}       
\usepackage{url}            
\usepackage{booktabs}       
\usepackage{amsfonts}       
\usepackage{microtype}      
\usepackage{amssymb,amsmath,amsthm,graphicx}
\usepackage{xcolor}
\usepackage{comment}
\usepackage{multirow}
\usepackage[caption=false]{subfig}
\usepackage{tabularx}
\usepackage{float}
\usepackage{algorithm}
\usepackage{algpseudocode}
\usepackage{booktabs}
\usepackage{enumitem} 
\usepackage{lipsum}
\usepackage{physics}
\usepackage[pagewise]{lineno}

\algrenewcommand\algorithmicrequire{\textbf{Input:}}
\algrenewcommand\algorithmicensure{\textbf{Output:}}
